\documentclass[prx,reprint,twocolumn,footinbib,longbibliography,superscriptaddress]{revtex4-2}
\usepackage{amsmath}
\usepackage{amsfonts}
\usepackage{amssymb}
\usepackage{lmodern}
\usepackage{graphicx}
\usepackage[usenames,dvipsnames]{xcolor}
\usepackage{bm}

\usepackage{physics}
\usepackage{mathrsfs}

\usepackage{hyperref}

\newcommand{\x}{\mathrm{x}}

\newcommand{\xy}{\mathrm{xy}}
\newcommand{\X}{\mathrm{X}}
\newcommand{\Y}{\mathrm{Y}}

\newcommand{\Z}{\mathrm{Z}}
\newcommand{\eq}{\mathrm{eq}}
\newcommand{\I}{\mathcal{I}}
\newcommand{\G}{\mathcal{G}}
\renewcommand{\u}{\mathrm{u}}

\renewcommand{\ss}{\text{ss}}
\newcommand{\st}{\mathsf{S}}
\newcommand{\gev}[1]{\left\langle\hspace{-2pt}\left\langle#1\right\rangle\hspace{-2pt}\right\rangle}

\begin{document}

\title{Information-Geometric Inequalities of Chemical Thermodynamics}
\date{\today}

\author{Kohei Yoshimura}
\affiliation{Department of Physics, The University of Tokyo, 7-3-1 Hongo, Bunkyo-ku, Tokyo 113-0031, Japan}
\author{Sosuke Ito}
\affiliation{Department of Physics, The University of Tokyo, 7-3-1 Hongo, Bunkyo-ku, Tokyo 113-0031, Japan}
\affiliation{JST, PRESTO, 4-1-8 Honcho, Kawaguchi, Saitama, 332-0012, Japan}

\begin{abstract}
    We study a connection between chemical thermodynamics and information geometry. We clarify a relation between the Gibbs free energy of an ideal dilute solution and an information-geometric quantity called an $f$-divergence. From this relation, we derive information-geometric inequalities that give a speed limit for a changing rate of the Gibbs free energy and a general bound of chemical fluctuations. 
    These information-geometric inequalities can be regarded as generalizations of the Cram\'{e}r--Rao inequality for chemical reaction networks described by rate equations, where unnormalized concentration distributions are of importance rather than probability distributions. 
    They hold true for damped oscillatory reaction networks and systems where the total concentration is not conserved so that the distribution cannot be normalized.
    We also formulate a trade-off relation between speed and time on a manifold of concentration distribution by using the geometrical structure induced by the $f$-divergence. 
    Our results apply to both closed and open chemical reaction networks, thus they are widely useful for thermodynamic analysis of chemical systems from the viewpoint of information geometry. 
\end{abstract}

\maketitle

\section{Introduction}
The history of chemical thermodynamics originates around the middle of the 19th century~\cite{gibbs1878ontheequilibrium, Kondepudi2014}. The chemical reaction in an ideal dilute solution is one of the main subjects in chemical thermodynamics. For example, the static nature of an ideal dilute solution under near-equilibrium condition has been well studied since then. After the invention of mathematics called chemical reaction network theory (CRNT) around the 1970s~\cite{horn1972general,feinberg1972complex}, its dynamic properties have also been well investigated~\cite{feinberg2019foundations}. One of the most important results of CRNT is that a class of chemical reaction networks called a complex balanced network has a Lyapunov function, which can be associated with thermodynamic quantities such as the Gibbs free energy~\cite{horn1972general,shear1967ananalog,higgins1968some}. 
This Lyapunov function is called the pseudo-Helmholtz function, and its time derivative is connected to the entropy production rate~\cite{ge2016nonequilibrium,rao2016nonequilibrium}. 

On the other hand, thermodynamics for stochastic processes have been well studied as stochastic thermodynamics~\cite{jarzynski1997nonequilibrium,sekimoto2010stochastic,seifert2012stochastic, Schmiedl2007stochastic}. In stochastic thermodynamics, physical quantities are given by probabilities, and we can discuss relations between thermodynamics and information theory~\cite{kawai2007dissipation, allahverdyan2009thermodynamic, sagawa2010generalized, toyabe2010experimental,still2012thermodynamics, sagawa2012fluctuation,ito2013information,horowitz2014thermodynamics,hartich2014stochastic,parrondo2015thermodynamics,ito2015maxwell,shiraishi2015fluctuation,rosinberg2016continuous} because probability plays a crucial role in information theory~\cite{cover2012elements}. For example, in recent years, stochastic thermodynamics met a branch of information theory called information geometry~\cite{amari2000methods,rao1945information,ruppeiner1995riemannian, crooks2007measuring,  rotskoff2015optimal,ito2018stochastic1, ito2018unified}, and its importance has been verified in recent studies of thermodynamic uncertainty relations~\cite{horowitz2019thermodynamic,barato2015thermodynamic,pietzonka2016universal,gingrich2016dissipation,polettini2016tightening,maes2017frenetic,horowitz2017proof,proesmans2017discrete-time,dechant2018current}. An information-geometric quantity called the Fisher information gives several geometric bounds such as the Cram\'{e}r--Rao bound~\cite{cover2012elements, rao1945information,amari2000methods} and these bounds indicate thermodynamic uncertainty relations in stochastic thermodynamics~\cite{ito2018stochastic1, dechant2018multidimensional,hasegawa2019uncertainty,ito2020stochastic2,otsubo2020estimation, ito2019glansdorff}.

Although analogy between stochastic thermodynamics and chemical thermodynamics has been studied~\cite{Schmiedl2007stochastic,ge2016nonequilibrium,rao2016nonequilibrium, rao2018conservation,falasco2019negative,avanzini2019thermodynamics,penocchio2019thermodynamic,wachtel2018thermodynamically,lazarescu2019large,ge2016nonequilibrium,ge2010physical,ge2012stochastic,peng2019universal}, connections between chemical thermodynamics and information theory are still vague because rate equations which govern chemical reactions are based on unnormalized concentration distributions rather than probability distributions. Nevertheless, a few researches have been conducted from the perspective of a connection between chemical thermodynamics and information theory~\cite{rao2016nonequilibrium,falasco2018information}, focusing on the fact that the pseudo-Helmholtz function has a similar form to the Kullback--Leibler divergence, which plays a fundamental role in information theory.

In this paper, we clarify a connection between chemical thermodynamics and information theory from the viewpoint of information geometry. In information geometry, $f$-divergence is well studied as a measure of the difference between two positive measures. The two positive measures do not have to be necessarily normalized like probability distributions and the Kullback--Leibler divergence may not be well defined for them. We show that the pseudo-Helmholtz function is not given by the Kullback--Leibler divergence, but by an $f$-divergence. Because the pseudo-Helmholtz function is a representative quantity of a chemical reaction system, this connection reveals how an information-geometric concept plays a fundamental role in chemical thermodynamics. Introducing a generalization of the Fisher information from an $f$-divergence and an average-like quantity which is more suitable to chemical reaction networks than the ordinary average, we obtain a generalization of the Cram\'{e}r--Rao inequality to CRNT. This generalized Cram\'{e}r--Rao inequality indicates the speed limit for the changing rate of the Gibbs free energy in terms of the fluctuation of the chemical potential. We numerically confirm the inequalities for a damped oscillatory reaction network, specifically the Brusselator model~\cite{prigogine1968symmetry}, and a system where the sum of concentrations is not conserved so that the distribution cannot be normalized.
We also examine the geometry of concentration distributions. We formulate trade-off relations between time and speed in terms of information geometry. The trade-off relations are numerically illustrated by solving an association reaction.

This paper is organized as follows. In Sec.~\ref{SecClosed}, we formulate chemical reaction networks and introduce the pseudo-Helmholtz function. Sec.~\ref{open} extends the formulation to open CRNs. Sec.~\ref{InfoGeo} is an introduction to information geometry of both probability spaces and positive measure spaces. We discuss mathematical properties of an $f$-divergence and the connection between the Fisher information and an $f$-divergence.
Sec.~\ref{sec_result} gives information-geometric inequalities, which are the main results of this paper. We show that the speed limits for the changing rate of the Gibbs free energy and more general observables are given by the intrinsic speed, the Fisher information, for both closed and open cases. Also, we indicate the speed limit is regarded as a generalization of the Cram\'{e}r--Rao inequality. 
In addition, we examine the geometrical structure of concentrations, and reveal trade-off relations between speed and time.
In Sec.~\ref{sec_example}, we confirm our main results through three characteristic models of chemical reaction networks. A conclusion and a further perspective of researches in Sec.~\ref{sec_conclusion}. 

\section{Thermodynamics of closed chemical reaction networks}
\label{SecClosed}
\subsection{Kinetics of chemical reaction networks}
In this paper, we consider the thermodynamics of a dilute solution with the temperature and pressure kept constant. Since the solvent is dominant, the volume is regarded as a constant. In this chapter, we focus on closed systems. 

We consider a chemical reaction network (CRN) consisting of $N$ species of molecules $\{\X_i\}_{i=1,2,\dots,N}$ in a closed vessel. A CRN is defined as a set of $M$ reactions
\begin{align}
    \sum_{i=1}^N\nu_{i\rho} \X_i
    \underset{k^-_{\rho}}{\overset{k^{+}_{\rho}}{\rightleftharpoons}}
    \sum_{i=1}^N\kappa_{i\rho} \X_i,
\end{align}
where reactions are labelled with $\rho=1,2,\dots,M$, stoichiometric coefficients $\nu_{i\rho}$, $\kappa_{i\rho}$ are nonnegative integers, and $k^\pm_\rho$ are rate constants. The reactions are assumed to be reversible. In the closed CRN, the time evolution of $\X_i$'s concentration $[\X_i]$ is described by the following rate equation
\begin{align}
    \frac{d[\X_i]}{dt}=\sum_{\rho=1}^M(\kappa_{i\rho}-\nu_{i\rho})J_\rho,\label{rateeq}
\end{align}
where $J_\rho$ is the reaction rate of the $\rho$-th reaction.
According to Waage--Guldberg's law of mass-action~\cite{lund1965guldberg}, the reaction rate $J_\rho$ is given by
\begin{gather}
    J_\rho=J_\rho^+-J_\rho^-,\\
    \text{with }
    J^{+}_{\rho}=k^{+}_{\rho}\prod_{i=1}^N [\X_i]^{\nu_{i\rho}},\;
    J^{-}_{\rho}=k^{-}_{\rho}\prod_{i=1}^N [\X_i]^{\kappa_{i\rho}},
\end{gather}
where $J_\rho^\pm$ is the forward/reverse reaction rate.
By the coefficients in Eq. \eqref{rateeq}, we define the stoichiometric matrix of the CRN as a $N\times M$ matrix with its $(i,\rho)$-element $\st_{i\rho}:=\kappa_{i\rho}-\nu_{i\rho}$, which corresponds to the change of $[\X_i]$ when one unit of $\rho$-th reaction proceeds. 
The rate equation can be written briefly in vector notation as
\begin{align}
    \frac{d[\bm{\X}]}{dt}=\st\bm{J}.
\end{align}

The rate equation Eq.~\eqref{rateeq} has \textit{a priori} conserved quantities. If $\bm{\ell}\in\mathbb{R}^N$ satisfies 
\begin{align}
    \bm{\ell}^\mathsf{T}\st=\bm{0}^\mathsf{T}, 
\end{align}
i.e., $\bm{\ell}\in\ker\st^\mathsf{T}=\{\bm{v}\mid\st^\mathsf{T}\bm{v}=\bm{0}\}$, the time derivative of $\bm{\ell}\cdot[\bm{\X}]$ is zero
\begin{align}
    \frac{d}{dt}(\bm{\ell}\cdot[\bm{\X}])=\bm{\ell}^\mathsf{T}\st\bm{J}=0,
\end{align}
where the superscript ${}^\mathsf{T}$ means transposition. 
Thus $\bm{\ell}\cdot[\bm{\X}]$ is conserved. 
We call a left null vector of a stoichiometric matrix a conservation law. Note that a closed CRN has at least one conservation law, corresponding to the conservation of the total mass. 

\subsection{Equilibrium conditions}

In thermodynamics, it is postulated that a closed system relaxes to equilibrium, at which a certain free energy is minimized depending on the condition. In the present case, the function to be minimized is the Gibbs free energy, thus the equilibrium distribution $[\bm{\X}]^\eq$ is defined as a distribution that minimizes the Gibbs free energy. 

Since the solution is supposed to be dilute, the Gibbs free energy per unit volume $G$ and the chemical potentials $\mu_i$ are expressed as
\begin{align}
    G&=\sum_{i=1}^N \mu_i[\X_i] -RT\sum_{i=1}^N[\X_i]+G_0,\label{Gibbs1}\\
    \mu_i&=\pdv{G}{[\X_i]}=\mu_i^\circ(T)+RT\ln[\X_i],
\end{align}
where $G_0$ is a constant, $R$ is the gas constant, and $\mu_i^\circ$ are the standard chemical potentials, which are independent of the concentration~\cite{Kondepudi2014,ge2016nonequilibrium,rao2016nonequilibrium}. It is plausible to call $G$ the Gibbs free energy because the volume does not change.

Possible value of the concentration is restricted because the concentration changes obeying the rate equation. That can be seen by integrating the rate equation. Let $\bm{s}_\rho$ be the $\rho$-th column vector of the stoichiometric matrix, $\st=(\bm{s}_1,\dots,\bm{s}_M)$. Then $[\bm{\X}]$ at $t$ is obtained as 
\begin{align}
    [\bm{\X}]
    &=[\bm{\X}]_0+\st\int_0^tdt\;\bm{J}\\
    &=[\bm{\X}]_0+\sum_{\rho=1}^M \pqty{\int_0^tdt\;J_\rho}\bm{s}_\rho,
\end{align}
where $[\bm{\X}]_0$ is the initial concentration. Thus a change of concentration $[\bm{\X}]-[\bm{\X}]_0$ must be a linear combination of $\{\bm{s}_\rho\}_{\rho=1,2,\dots,M}$. 
So a set of concentrations that $[\bm{\X}]$ may reach is given by 
\begin{align}
    \mathcal{S}([\bm{\X}]_0):=\{[\bm{\X}]_0+\st\bm{\xi} \mid \bm{\xi} \in \mathbb{R}^M\} \cap \mathbb{R}_{\geq0}^N, 
\end{align}
where $\mathbb{R}_{\geq0}^N$ is the set of $N$-dimensional vectors with nonnegative elements. This set is called the stoichiometric compatibility class~\cite{feinberg2019foundations}, and $\bm{\xi}$ the extent of reaction. 

A necessary condition for equilibrium can be obtained as that the derivative of the Gibbs free energy with respect to $\xi_\rho$ vanishes for all $\rho$, 
\begin{align}
    \pdv{G}{\xi_\rho}=\sum_{i=1}^N\pdv{[\X_i]}{\xi_\rho}\pdv{G}{[\X_i]}
    =\sum_{i=1}^N\st_{i\rho}\mu_i^\eq
    =((\bm{\mu}^\eq)^\mathsf{T}\st)_\rho=0.
    \label{eqcond}
\end{align}
So the equilibrium distribution $[\bm{\X}]^\eq$ satisfies
\begin{align}
    \sum_{i=1}^N(\mu_i^\circ+RT\ln[\X_i]^\eq)\st_{i\rho}=0. 
    \label{17}
\end{align}
Note that it means that the equilibrium chemical potential $\bm{\mu}^\eq$ is a conservation law. 

On the other hand, equilibrium state is often characterized by the detailed balance, which is based on kinetics, 
\begin{align}
    J^+_\rho=J^-_\rho. \label{db}
\end{align}
The consistency between the thermodynamic condition of equilibrium \eqref{eqcond} and this detailed balance condition leads to a relation 
\begin{align}
    \frac{k^{+}_{\rho}}{k^{-}_{\rho}}
    =\exp(-\frac{({\bm{\mu}^\circ}^\mathsf{T}\st)_\rho}{RT}),\label{localDB}
\end{align}
which is called the local detailed balance property. It can be regarded as a bridge between thermodynamics and kinetics.

\subsection{Gibbs free energy with divergence}\label{GEcl}
\begin{figure}
    \begin{center}
        \includegraphics[width=\hsize]{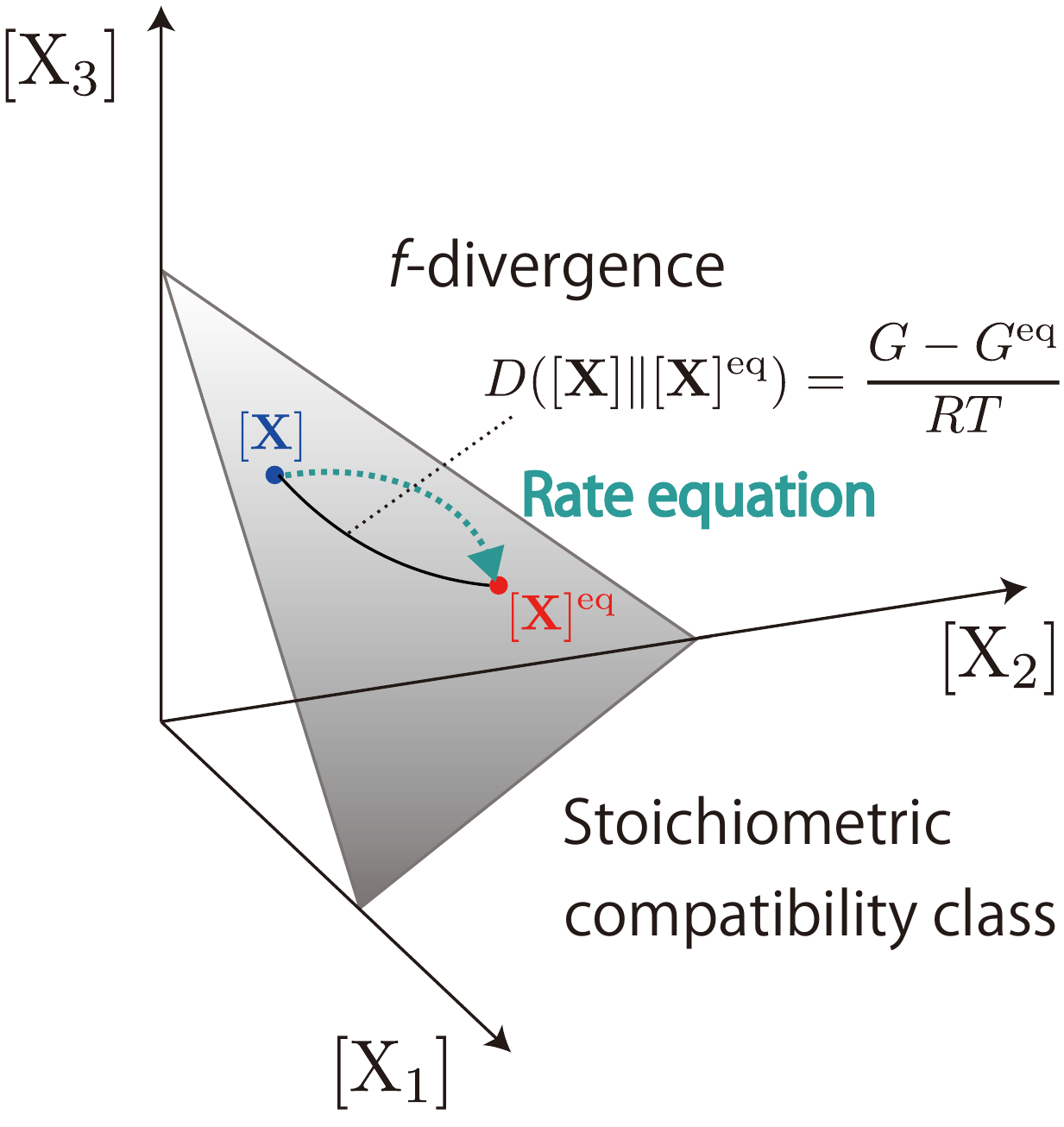}
        \caption{Schematic of a stoichiometric compatibility class and the $f$-divergence for $N=3$. This $f$-divergence gives the difference of the Gibbs free energy $(G-G^\eq)/RT$.}
        \label{divzu}
    \end{center}
\end{figure}
From Eq.~\eqref{eqcond}, $\bm{\mu}^\eq$ is a conservation law, therefore $\bm{\mu}^\eq\cdot[\bm{\X}]$ is time invariant 
\begin{align}
    \frac{d}{dt}\pqty{\bm{\mu}^\eq\cdot[\bm{\X}]}=(\bm{\mu}^\eq)^\mathsf{T}\st\bm{J}=0. \label{conservedmu}
\end{align}
So the Gibbs free energy at equilibrium $G^\eq$ can be expressed as 
\begin{align}
    G^\eq&=\bm{\mu}^\eq \cdot[\bm{\X}]^\eq-RT\sum_{i=1}^N[\X_i]^\eq+G_0\\
    &=\bm{\mu}^\eq\cdot[\bm{\X}]-RT\sum_{i=1}^N[\X_i]^\eq+G_0
\end{align}
for a concentration $[\bm{\X}](\neq[\bm{\X}]^\eq)$ of an arbitrary time, namely, in the same stoichiometric compatibility class as $[\bm{\X}]^\eq$. As a result, we obtain the expression 
\begin{align}
    &G-G^\eq\notag\\
    &=(\bm{\mu}-\bm{\mu}^\eq)\cdot[\bm{\X}]-RT\sum_{i=1}^N[\X_i]
    +RT\sum_{i=1}^N[\X_i]^\eq\\
    &=RT\sum_{i=1}^N\pqty{[\X_i]\ln\frac{[\X_i]}{[\X_i]^\eq}-[\X_i]+[\X_i]^\eq}, \label{Gibbs2}
\end{align}
by which a function is usually defined as 
\begin{align}
    D([\bm{\X}]\|[\bm{\X}]^\eq):=\sum_{i=1}^N\pqty{[\X_i]\ln\frac{[\X_i]}{[\X_i]^\eq}-[\X_i]+[\X_i]^\eq}. \label{Div_shoshutsu}
\end{align}
This suggestive form has been known for more than a half century, and $D([\bm{\X}]\|[\bm{\X}]^\eq)$ called the pseudo-Helmholtz function~\cite{horn1972general}, Shear's Lyapunov function~\cite{shear1967ananalog,higgins1968some} or the relative entropy~\cite{rao2016nonequilibrium} (see Fig.~\ref{divzu}). 
In terms of information geometry~\cite{amari2000methods}, it is regarded as an $f$-divergence of a positive measure space. As we see in Sec. \ref{InfoGeo}, an $f$-divergence is nonnegative and equal to zero if and only if the two arguments coincide. Therefore $G$ is greater than or equal to $G^\eq$, and $G=G^\eq$ only in equilibrium, $[\bm{\X}]=[\bm{\X}]^\eq$. 

It is also known that the Gibbs free energy of a closed system never increases under the mass-action kinetics, i.e., it is a Lyapunov function of a closed CRN. To show this fact, we calculate the time derivative of the Gibbs free energy
\begin{align}
    \frac{dG}{dt}=\bm{\mu}\cdot\frac{d[\bm{\X}]}{dt}
    =\bm{\mu}^\mathsf{T}\st\bm{J}. \label{eq22}
\end{align}
From the local detailed balance property Eq. \eqref{localDB}, $\bm{\mu}^\mathsf{T}\st$ can be transformed as
\begin{align}
    (\bm{\mu}^\mathsf{T}\st)_\rho
    &=({\bm{\mu}^\circ}^\mathsf{T}\st)_\rho+RT\ln\prod_{i=1}^N[\X_i]^{\st_{i\rho}} \notag\\
    &=-RT\ln\frac{k^+_\rho\prod_{i=1}^N[\X_i]^{\nu_{i\rho}}}
    {k^-_\rho\prod_{i=1}^N[\X_i]^{\kappa_{i\rho}}} \notag\\
    &=-RT\ln\frac{J^+_\rho}{J^-_\rho}.\label{potentialtocurrent}
\end{align}
By substituting it into Eq.~\eqref{eq22}, we have
\begin{align}
    \frac{dG}{dt}=-RT\sum_{\rho=1}^M(J^+_\rho-J^-_\rho)\ln\frac{J^+_\rho}{J^-_\rho}\leq 0\label{eq25}.
\end{align}
The last inequality follows from the fact that the signs of $J^+_\rho-J^-_\rho$ and $\ln(J^+_\rho/J^-_\rho)$ are always the same for all $\rho$. Since $dG/dt$ is negative unless the detailed balance is satisfied, and $G$ is always not less than $G^\eq$, hence $G$ decreases to $G^\eq$ monotonically. 
Here, we point out two facts. First, the left-hand side of the inequality Eq.~\eqref{eq25} coincides with the opposite sign of the entropy production rate, which we define in the next section. Thus the inequality expresses the second law of thermodynamics. Second, the left-hand side is also represented by the $f$-divergence $D$ between $\bm{J}^+$ and $\bm{J}^-$ 
\begin{align}
    \frac{dG}{dt}=-RT\pqty{D(\bm{J}^+\|\bm{J}^-)+D(\bm{J}^-\|\bm{J}^+)}.
\end{align}

\subsection{Entropy production rate and affinity}
We can formulate the second law of thermodynamics in CRN~\cite{rao2016nonequilibrium}. Here we review a few important points. The entropy production rate $\sigma$ due to chemical reactions is given by 
\begin{align}
    \sigma=R\sum_{\rho=1}^M(J_\rho^+-J_\rho^-)\ln\frac{J_\rho^+}{J_\rho^-}. \label{epr}
\end{align}
So it coincides with $-(1/T)dG/dt$ in closed CRNs. 
As we explained in the preceding section, the entropy production rate is always positive except for the equilibrium state, and that indicates the second law of thermodynamics. 

The affinity $F_\rho$ of a reaction is defined as follows~\cite{deDonder1936thermodynamic} 
\begin{align}
    F_\rho:=\sum_{i=1}^N\nu_{i\rho}\mu_i-\sum_{i=1}^N\kappa_{i\rho}\mu_i=-(\bm{\mu}^\mathsf{T}\st)_{\rho}.
\end{align}
The affinity corresponds to the energy difference between the reactant and product of the $\rho$-th reaction. From the local detailed balance property Eq.~\eqref{localDB}, one can rewrite it as 
\begin{align}
    F_\rho=RT\ln\frac{J_\rho^+}{J_\rho^-}. 
\end{align}
Thus, the entropy production is expressed as the sum of the product between the reaction rate and the affinity 
\begin{align}
    T\sigma=\sum_{\rho=1}^MJ_\rho F_\rho. 
\end{align}
This expression shows there is an analogy with stochastic thermodynamics of master equations~\cite{schnakenberg1976network}. 

\section{Themodynamics of open chemical reaction networks}\label{open}

We have already seen the connection between the Gibbs free energy $G$ and the $f$-divergence $D$ in closed CRNs. On the other hand, in open CRNs, the Gibbs free energy is not directly related to the $f$-divergence, even in a CRN called a complex balanced network where the $f$-divergence becomes a Lyapunov function. This is because the chemical potential at a steady state is not a conservation law in general. However, we can associate the Gibbs free energy with the $f$-divergence. 
In this chapter, we formulate open CRNs and associate the Gibbs free energy with the $f$-divergence in two ways, the method already known and the one we newly propose, respectively. 

\subsection{Setup for open CRNs}
Let $\{\Y_j\}_{j=1,2,\dots,N'}$ be the chemical species that are exchanged with the environment, and $\{\X_i\}_{i=1,2,\dots,N}$ be the other internal species. The former is assumed to be chemostatted, that is, their concentrations are constant. 

The CRN consisting of them is expressed as 
\begin{align}
    \sum_{i=1}^N\nu_{i\rho} \X_i
    &+\sum_{j=1}^{N'} \nu_{(N+j)\rho}\Y_j\notag\\
    &\underset{k^-_{\rho}}{\overset{k^{+}_{\rho}}{\rightleftharpoons}}
    \sum_{i=1}^N\kappa_{i\rho} \X_i
    +\sum_{j=1}^{N'} \kappa_{(N+j)\rho}\Y_j.
\end{align}
The stoichiometric matrix $\st=\qty(\kappa_{i\rho}-\nu_{i\rho})_{i=1,\dots,N+N'}^{\rho=1,\dots,M}$ can be decomposed into an $N$ rows of $\X$ part and an $N'$ rows of $\Y$ part 
\begin{align}
    \st=\pmqty{\st^\X \\ \st^\Y},  
\end{align}
where $\st^\X_{i\rho}=\kappa_{i\rho}-\nu_{i\rho}$ $(i=1,2,\dots,N)$ and $\st^\Y_{j\rho}=\kappa_{(j+N)\rho}-\nu_{(j+N)\rho}$ $(j=1,2,\dots,N')$. 
Hereafter, for an $N+N'$ row quantity $\mathsf{Q}$, let $\mathsf{Q}^\X$ be the first $N$ rows and $\mathsf{Q}^\Y$ the remainder as the above case. 
Since the concentrations of the chemostatted species are assumed to be constant, the dynamics are expressed by the following rate equation 
\begin{align}
    \frac{d[\bm{\Z}]}{dt}:=\frac{d}{dt}\pmqty{[\bm{\X}] \\ [\bm{\Y}]}
    =\pmqty{\st^\X\bm{J} \\ \bm{0}},
\end{align}
where $[\bm{\Z}]^\mathsf{T}=\pmqty{[\bm{\X}]^\mathsf{T} & [\bm{\Y}]^\mathsf{T}}$, $J_\rho=J_\rho^+-J_\rho^-$ and 
\begin{align}
    J_\rho^+&=k_\rho^+\prod_{i=1}^N[\X_i]^{\nu_{i\rho}}\prod_{j=1}^M[\Y_j]^{\nu_{(j+N)\rho}}\\
    J_\rho^-&=k_\rho^-\prod_{i=1}^N[\X_i]^{\kappa_{i\rho}}\prod_{j=1}^M[\Y_j]^{\kappa_{(j+N)\rho}}.
\end{align}
A steady state is defined as a state at which the concentration does not change in time, $\st^\X\bm{J}=\bm{0}$.

In open CRNs, $\bm{\ell}\cdot[\bm{\Z}]$ is not necessarily conserved even if $\bm{\ell}$ is a conservation law, i.e., $\bm{\ell}^\mathsf{T}\st=\bm{0}^\mathsf{T}$ holds. Since we have 
\begin{align}
    \frac{d}{dt}(\bm{\ell}\cdot[\bm{\Z}])=(\bm{\ell}^\X)^\mathsf{T}\st^\X\bm{J}, 
\end{align}
$\bm{\ell}\cdot[\bm{\Z}]$ is conserved if $\bm{\ell}^\X$ belongs to $\ker(\st^\X)^\mathsf{T}$. Thus a conservation law $\bm{\ell}$ leads to a conserved quantity if $\bm{\ell}$ is an element of the linear space $\pqty{\ker(\st^\X)^\mathsf{T}\times \mathbb{R}^{N'}}\cap\ker\st^\mathsf{T}=:L_\X$, where $\times$ means the direct product between two linear spaces. 

\subsection{One way to associate $G$ with $D$}
One way to associate the Gibbs free energy with the $f$-divergence is to decompose a basis of the space of conservation laws $\ker\st^\mathsf{T}$. This formalism is based on Rao and Esposito's paper Ref.~\cite{rao2016nonequilibrium}. It can be used only when the steady state is detailed balanced. 

To construct the desired basis, we exploit the linear space $L_\X$. Let $\{\bm{\ell}^{\lambda_\x}\}_{\lambda_\x=1,2,\dots,\Lambda_\x}$ be a basis of $L_\X$, then we obtain a basis of $\ker\st^\mathsf{T}$ by adding some vectors $\{\bm{\ell}^{\lambda_\xy}\}_{\lambda_\xy=1,2,\dots,\Lambda_\xy}$. The former vectors lead to quantities $\bm{\ell}^{\lambda_\x}\cdot[\bm{\Z}]$ that are conserved in an open CRN. We call them X-conservation laws and the latter XY-conservation laws. XY-conservation laws $\bm{\ell}^{\lambda_\xy}$ are usually called broken laws because the remainder, X-conservation laws, are always true conservation laws. However, since an XY-conservation law might be a true conservation law, we do not use the conventional terminology. 

From the local detailed balanced condition Eq.~\eqref{localDB}, we have the following relation (see Eq.~\eqref{potentialtocurrent}) 
\begin{align}
    (\bm{\mu}^\mathsf{T}\st)_\rho=-RT\ln\frac{J^+_\rho}{J^-_\rho}. 
\end{align}
So if we suppose the steady state to be detailed balanced $J_\rho^+=J_\rho^-$, i.e., be an equilibrium, the chemical potential at equilibrium $\bm{\mu}^\eq$ becomes a conservation law. Then we can expand it with the prepared basis as 
\begin{align}
    \bm{\mu}^\eq=\sum_{\lambda_\x=1}^{\Lambda_\x}f_{\lambda_\x}\bm{\ell}^{\lambda_\x}
    +\sum_{\lambda_\xy=1}^{\Lambda_\xy}f_{\lambda_\xy}\bm{\ell}^{\lambda_\xy}.
    \label{expmu}
\end{align}

A function $\G_1$ is defined by 
\begin{align}
    \G_1:=G-\sum_{\lambda_\xy=1}^{\Lambda_\xy}f_{\lambda_\xy}\bm{\ell}^{\lambda_\xy}\cdot[\bm{\Z}], \label{defg1}
\end{align}
which is called the transformed Gibbs free energy in Ref.~\cite{rao2016nonequilibrium}. 
From Eq.~\eqref{Gibbs1} and the expansion of $\bm{\mu}^\eq$ Eq.~\eqref{expmu}, we see 
\begin{equation}
    \begin{split}
        \G_1=\sum_{i=1}^{N+N'}([\Z_i]\mu_i&-RT[\Z_i])+G_0\\
        &-\pqty{\bm{\mu}^\eq\cdot[\bm{\Z}]-\sum_{\lambda_\u=1}^{\Lambda_\u}f_{\lambda_\u}\bm{\ell}^{\lambda_\u}\cdot[\bm{\X}]}.
    \end{split}
\end{equation}
Since the concentrations of the chemostatted species are constant, $[\Y_j]$ coincide with $[\Y_j]^\eq$. Therefore we have 
\begin{align}
    \G_1&=\sum_{i=1}^N\pqty{[\X_i](\mu_i-\mu_i^\eq)-RT[\X_i]}
    +\text{const.}\\
    &=RT\sum_{i=1}^N\pqty{[\X_i]\ln\frac{[\X_i]}{[\X_i]^\eq}-[\X_i]+[\X_i]^\eq}
    +\text{const.}, 
\end{align}
where we use the fact that $RT[\bm{\X}]^\eq$ is constant. Since the last constant term is equal to $\G_1$ at the equilibrium, we write it $\G_1^\eq$. We finally obtain the following equation as in the closed CRN 
\begin{align}
    \G_1=\G_1^\eq+RTD([\bm{\X}]\|[\bm{\X}]^\eq). 
\end{align}
As shown in Ref.~\cite{rao2016nonequilibrium}, $\G_1$ gives a bound to the irreversible work to manipulate nonequilibrium distributions. Its time derivative also provides the non-adiabatic entropy production rate as we will see later. The arbitrariness about the choice of the basis yields only a constant term~\cite{rao2018conservation}. However, we note that it is only defined for CRNs that relax to detailed balanced steady states. Therefore, if one uses $\G_1$, the number of systems that can be examined with the $f$-divergence, or in other words, the relative entropy, would be limited.

\subsection{Another way to associate $G$ with $D$}
We newly propose another way of association, which is simpler and more widely applicable than the preceding one. Letting $[\bm{\Z}]^\ss$ be the concentration at a steady state and $\bm{\mu}^\ss=(\mu_i^\circ+RT\ln[\Z_i]^\ss)_{i=1,\dots,N+N'}$ the chemical potential, we define 
\begin{align}
    \G_2:=G-\bm{\mu}^\ss\cdot[\bm{\Z}]. \label{tg2}
\end{align}
Since $[\Y_j]$ coincide with $[\Y_j]^\ss$ as in the detailed balanced case, we have 
\begin{align}
    \G_2&=\sum_{i=1}^N([\X_i](\mu_i-\mu_i^\ss)-RT[\X_i])+\text{const.}\\
    &=\sum_{i=1}^N([\X_i](\mu_i-\mu_i^\ss)-RT[\X_i]+RT[\X_i]^\ss)+\G_2^\ss\\
    &=\G_2^\ss+RTD([\bm{\X}]\|[\bm{\X}]^\ss), 
\end{align}
where $\G_2^\ss$ is $\G_2$ at the steady state. 

$\G_2$ is defined for general steady states and coincides with $\G_1$ up to a constant if the steady state is detailed balanced. In fact, 
\begin{align}
    \G_2&=G-\bm{\mu}^\eq\cdot[\bm{\Z}]\\
    &=\G_1-\sum_{\lambda_\x=1}^{\Lambda_\x}f_{\lambda_\x}\bm{\ell}^{\lambda_\x}\cdot[\bm{\Z}]
\end{align}
and $\bm{\ell}^{\lambda_\x}\cdot[\bm{\Z}]$ are constants, so $\G_2-\G_1=\mathrm{const}$. 
Thus $\G_2$ is a generalization of $\G_1$, then we denote $\G_2$ as $\G$ and call it the transformed Gibbs free energy. 

The transformed Gibbs free energy we introduce here can be defined for open CRNs that do not satisfy detailed balance. Not all CRNs have steady-state solutions, but our definition Eq.~\eqref{tg2} enables us to study much broader class of open CRNs than the previous one Eq.~\eqref{defg1} does.
$\G_2$ also has the merit that it removes the arbitrariness in the choice of the basis as $\G_1$.
For open CRNs with multiple steady states, one can define the transformed Gibbs free energy by choosing a steady state in Eq.~\eqref{tg2}. 
Remarkably, the following results hold regardless of the choice of the steady state.
We comment that CRNs that do not have steady-state solutions are outside our framework. They include open CRNs that sustain oscillations. 

\subsection{Entropy production rate and affinity}
The entropy production rate $\sigma$ and the affinity $F_\rho$ of an open CRN have the same form as of closed one. The entropy production rate of an open CRN can be decomposed into the adiabatic and non-adiabatic parts $\sigma_\mathrm{a}$, $\sigma_\mathrm{na}$~\cite{ge2016nonequilibrium,rao2016nonequilibrium} which are defined as 
\begin{align}
    \sigma&=\sigma_{\mathrm{a}}+\sigma_{\mathrm{na}}\\
    \sigma_{\mathrm{a}}&:=R\sum_{\rho=1}^MJ_\rho\ln \frac{J_\rho^{+,\ss}}{J_\rho^{-,\ss}}\\
    \sigma_{\mathrm{na}}&:=R\sum_{\rho=1}^MJ_\rho\ln \frac{J_\rho^+J_\rho^{-,\ss}}{J_\rho^-J_\rho^{+,\ss}}, 
\end{align}
where $J_\rho^{\pm,\ss}$ are the reaction rates at the steady state. 
The time derivative of the transformed Gibbs free energy gives the minus sign of the non-adiabatic entropy production rate $d\G/dt=-T\sigma_{\mathrm{na}}$. If the steady state is detailed balanced, or equivalently, $J_\rho^{+,\ss}=J_\rho^{-,\ss}$ holds, $d\G/dt$ can be written by the $f$-divergence between the reaction rates as in the closed case 
\begin{align}
    -\frac{d\G}{dt}
    =RT(D(\bm{J}^+\|\bm{J}^-)+D(\bm{J}^-\|\bm{J}^+)). 
\end{align}

\section{Information Geometry} \label{InfoGeo}
Information geometry deals with a manifold of probability distributions $\bm{p}=(p_i)_{i=1,2,\dots,N}\in\mathbb{R}^N_{>0}$ that satisfy the normalization condition $\sum_{i=1}^Np_i=1$, or a manifold of positive measures on a discrete set $\bm{m}=(m_i)_{i=1,2,\dots,N}\in\mathbb{R}^N_{>0}$, which does not have to be normalized~\cite{amari2000methods}. The former manifold is called a probability simplex, and the latter a positive measure space. We use the term "distribution" for either a probability distribution or a positive measure in this section. 

\subsection{$f$-divergence}\label{subsec_fdiv}
A divergence $D(\cdot\|\cdot)$ is a measure of the separation between two distributions $\bm{m}$ and $\bm{n}$ that satisfies the following conditions~\cite{amari2000methods}: 
\begin{enumerate}
    \item $D(\bm{m}\|\bm{n})\geq 0$
    \item $D(\bm{m}\|\bm{n})=0\iff \bm{m}=\bm{n}$
    \item $\displaystyle D(\bm{m}\|\bm{m}+d\bm{m})=\frac{1}{2}\sum_{i,j=1}^N g_{ij} dm_idm_j
    +o(dm^2)$, \\and the matrix $(g_{ij})_{\,1\leq i,j\leq N}$ is positive definite.
\end{enumerate}
Note that a divergence is similar to a distance function but it is not really because it is not symmetric $D(\bm{m}\|\bm{n})\neq D(\bm{n}\|\bm{m})$. 

One of well-known divergences is an $f$-divergence~\cite{csiszar1991least}, which has the following form 
\begin{align}
    D(\bm{m}\|\bm{n})=\sum_{i=1}^N m_if\pqty{\frac{n_i}{m_i}}. \label{fdiv}
\end{align}
The function $f$ has to fulfill some conditions. 
For both kinds of manifold, $f$ should be a convex differentiable function which satisfies $f(1)=0$. If one chooses $f(x)=-\ln x$, $D$ becomes the Kullback--Leibler divergence. For positive measure spaces, the condition $f'(1)=0$ is imposed additionally. $f$ is called a standard convex function when $f''(1)=1$ holds. 

The nonnegativity of an $f$-divergence is easily proved. Let $\bm{p}$ and $\bm{q}$ be probability distributions in a probability simplex. From Jensen's inequality, we have 
\begin{align}
    \sum_{i=1}^N p_if\pqty{\frac{q_i}{p_i}}\geq f\pqty{\sum_{i=1}^Np_i\frac{q_i}{p_i}}=f(1)=0. 
\end{align}
Therefore, an $f$-divergence on a probability simplex is nonnegative and equal to zero if and only if $\bm{p}=\bm{q}$. 
On the other hand, for a positive measure manifold, since $f$ is convex  and $f'(1)=0$, $f$ takes the minimum value $0$ at $x=1$. Thus $f$ is nonnegative, and so is an $f$-divergence since the all coefficients of $f$ are positive. An $f$-divergence is zero if and only if $f(n_i/m_i)=0$ for all $i$. It is equivalent to that two distributions are the same. 

\subsection{Fisher information}
Letting $\theta$ be the parameter of distributions and $D$ a divergence, 
the Fisher information $\I(\theta)$~\cite{amari2000methods} is defined as 
\begin{align}
    \I(\theta)&:=\sum_{i,j=1}^N g_{ij}\frac{dm_i}{d\theta}\frac{dm_j}{d\theta}, \\
    g_{ij}&=\eval{\pdv{m_i}\pdv{m_j}D(\bm{n}\|\bm{m})}_{\bm{n}=\bm{m}}.
\end{align}
In information geometry, 
\begin{align}
    ds^2:=\sum_{i,j=1}^N g_{ij} dm_idm_j\simeq 2D(\bm{m}\|\bm{m}+d\bm{m})
\end{align}
is interpreted as the square of the line element between two close distributions. If the distributions are parametrized by the time $t$, the Fisher information becomes 
\begin{align}
    \I(t)=\sum_{i,j=1}^N g_{ij}\frac{dm_i}{dt}\frac{dm_j}{dt}=\frac{ds^2}{dt^2}, 
\end{align}
so we define the intrinsic speed on the manifold $ds/dt$ as 
\begin{align}
    \frac{ds}{dt}:=\sqrt{\I(t)}.
\end{align}

An $f$-divergence with a standard convex function leads to $g_{ij}=m_i^{-1}\delta_{ij}$, where $\delta_{ij}$ is Kronecker's delta, then the Fisher information always has the unique form 
\begin{align}
    \I(\theta)=\sum_{i=1}^N\frac{1}{m_i}\pqty{\frac{dm_i}{d\theta}}^2. 
\end{align}
A significant fact related to the Fisher information is the Cram\'{e}r--Rao inequality for a probability distribution $\bm{p}(\theta)$~\cite{rao1945information} 
\begin{align}
    \mathrm{Var}(\hat{\theta})\geq \frac{1}{\I(\theta)}, \label{crineq}
\end{align}
where $\hat{\theta}$ is an unbiased estimator of $\theta$, that is, $\langle\hat{\theta}\rangle_\theta:=\sum_{i=1}^Np_i(\theta)\hat{\theta}_i=\theta$ holds, and $\mathrm{Var}(\hat{\theta})$ is the variance of $\hat{\theta}$, $\langle(\hat{\theta}-\langle\hat{\theta}\rangle_\theta)^2\rangle_\theta$.

\section{Information geometry in chemical thermodynamics}\label{sec_result}
\subsection{Geometrical structure of chemical thermodynamics}\label{sec_va}
A set of concentrations can be interpreted as a positive measure space in both closed and open CRNs. The measures are concentrations $[\bm{\Z}]$, and the measurable set is the index set of species. This space is thought to have the \textit{a priori} $f$-divergence, with its standard convex function $f(x)=-\ln x+x-1$, 
\begin{align}
    D([\bm{\X}]\|[\bm{\X}'])&=\sum_{i=1}^N [\X_i]f\pqty{\frac{[\X'_i]}{[\X_i]}} \notag\\
    &=\sum_{i=1}^N \pqty{[\X_i]\ln\frac{[\X_i]}{[\X'_i]}-[\X_i]+[\X'_i]}, \label{concdiv}
\end{align}
because $G$ of a closed CRN and $\mathcal{G}$ of an open CRN are described by the $f$-divergence as 
\begin{align}
    G=G^\eq+RTD([\bm{\X}]\|[\bm{\X}]^\eq) \label{GibbsDiv}
\end{align}
and
\begin{align}
    \G=\G^\ss+RTD([\bm{\X}]\|[\bm{\X}]^\ss). \label{transfGibbsDiv}
\end{align}
Let us confirm that $f$ is a standard convex function. It is obviously smooth, and convex because $f''(x)=1/x^2>0$. It is readily seen that the values at $x=1$ are $f(1)=f'(1)=0$ and $f''(1)=1$. Hence, $f$ is a standard convex function. 

As we pointed out in Sec.~\ref{InfoGeo}, a divergence gives a geometrical structure to a manifold, namely, a metric $g_{ij}$. An $f$-divergence always provides the metric of the form $g_{ij}=\delta_{ij} m_i^{-1}$. Therefore, manifolds of concentration distributions are considered to be equipped with the metric $g_{ij}=\delta_{ij} [\X_i]^{-1}$. The metric is given naturally with respect to only internal species. That is because the concentrations of the chemostatted species are kept constant and do not appear in the divergence Eq.~\eqref{transfGibbsDiv}. It is possible to extend the metric as $g_{ij}=\delta_{ij}[\Z_i]^{-1}$ for $i,j\in\{1,2,\dots,N+N'\}$, but we just neglect the chemostatted species in this paper when we consider the geometrical structure of CRNs. Hence, we only say here that the concentration distributions of closed and open CRNs have the completely same metric $g_{ij}=\delta_{ij}[\X_i]^{-1}$, where $i,j\in \{1,2,\dots,N\}$. 

Before stating our main results, we denote a few direct consequences of the geometrical structure. 
First, the square of the line element between the equilibrium concentration and a concentration close to it is related to the Gibbs free energy. If a concentration is close to equilibrium $[\bm{\X}]=[\bm{\X}]^\eq+\delta[\bm{\X}]$, the Gibbs free energy is expressed as $G=G^\eq+\delta G$ with a small deviation $\delta G$. From Eq.~\eqref{GibbsDiv}, we obtain the expression of the square of the line element $ds^2=2D([\bm{\X}]\|[\bm{\X}]^\eq)$ by the Gibbs free energy 
\begin{align}
    ds^2 = \frac{2}{RT}\delta G. \label{lineelem}
\end{align}
So the Gibbs free energy difference corresponds to the square of the distance from the equilibrium concentration under near-equilibrium condition.
It is true for open CRNs, that is, the fluctuation of the transformed Gibbs free energy from the steady-state value $\delta \G$ corresponds to $ds^2$ between the steady state and a concentration nearby $ds^2=(2/RT)\delta\G$. 
Note that although the relations hold only under near-equilibrium or near-steady-state condition, our results in the following subsections apply to far-from-equilibrium systems except for Eqs.~\eqref{fishneareq}--\eqref{fisherneareqderive}. 

Second, we can consider the Fisher information of chemical reaction networks. Since the metric is $\delta_{ij}/[\X_i]$ for closed and open CRNs and $d[\Y_j]/dt=0$, the Fisher information $\I(t)$ is defined for both types of CRN as 
\begin{align}
    \I(t)=\sum_{i=1}^N\frac{1}{[\X_i]}\pqty{\frac{d[\X_i]}{dt}}^2. 
\end{align}
It is also represented as 
\begin{align}
    \I(t)=-\frac{1}{RT}\sum_{\rho=1}^MJ_\rho\frac{dF_\rho}{dt}. \label{fishjf}
\end{align}
We prove this formula that is true in both cases. 
Because $(d/dt)\ln[\X_i]=(1/RT)(d\mu_i/dt)$, $\I(t)$ is transformed as follows 
\begin{align}
    \I(t)
    &=\frac{1}{RT}\sum_{i=1}^N\frac{d[\X_i]}{dt}\frac{d\mu_i}{dt}\\
    &=\frac{1}{RT}\sum_{i=1}^N\sum_{\rho=1}^M\frac{d\mu_i}{dt}\st_{i\rho}J_\rho.\\
    \intertext{If the CRN is closed, that ends the proof since $F_\rho=-(\bm{\mu}^\mathsf{T}\st)_\rho$. On the other hand, if open }
    \I(t)&=\frac{1}{RT}\sum_{i=1}^{N+N'}\sum_{\rho=1}^M\frac{d\mu_i}{dt}\st_{i\rho}J_\rho\\
    &=-\frac{1}{RT}\sum_{\rho=1}^M\frac{dF_\rho}{dt}J_\rho, 
\end{align}
where we used the fact that the chemical potentials of the chemostatted species do not change in time. 

From the expression of the entropy production rate Eq.~\eqref{epr}, which is valid for closed and open CRNs, Eq.~\eqref{fishjf} is also written as 
\begin{align}
    \I(t)=-\frac{1}{R}\frac{d\sigma}{dt}+\frac{1}{RT}\sum_{\rho=1}^M\frac{dJ_\rho}{dt}F_\rho. \label{fishjf2}
\end{align}
Therefore the Fisher information is associated with the entropy production rate with the additional term. Furthermore, under near-equilibrium conditions, $\I(t)$ is directly given by $d\sigma/dt$ with an error of higher order in $\delta J$. That is, 
\begin{align}
    \I(t)=-\frac{1}{2R}\frac{d\sigma}{dt} +o(\delta J^2) \label{fishneareq}
\end{align}
is obtained. This is because the affinity is proportional to the reaction rate under near-equilibrium conditions. It can be proved as follows. 
$J_\rho^\pm$ can be decomposed into the equilibrium values $J_\rho^{+,\eq}=J_\rho^{-,\eq}=:J_\rho^{\eq}$ and the fluctuations $\delta J_\rho^{\pm}$ as $J_\rho^{\pm}=J_\rho^\eq+\delta J_\rho^{\pm}$. Then the affinities are obtained as $F_\rho= RTJ_\rho/J_\rho^\eq+o(J_\rho)$, where $J_\rho=\delta J_\rho^+-\delta J_\rho^-$. So, we have 
\begin{align}
    \sum_{\rho=1}^M\frac{dJ_\rho}{dt}F_\rho
    &\simeq \sum_{\rho=1}^MRT\frac{J_\rho}{J_\rho^\eq}\frac{dJ_\rho}{dt}\\
    &\simeq \frac{1}{2}\frac{d}{dt}\pqty{\sum_{\rho=1}^M
    J_\rho F_\rho}
    =\frac{T}{2}\frac{d\sigma}{dt}. \label{fisherneareqderive}
\end{align}
By substituting it into Eq.~\eqref{fishjf2}, we obtain the equality Eq.~\eqref{fishneareq}.

\subsection{Speed limit on the Gibbs free  energy}\label{sec_speedlimit}
We describe the main result in this section as follows. We state, discuss, and prove our assertion for closed CRNs. Then we obtain the same result in subsystems and open CRNs. That is because the discussion becomes concise without loss of generality in closed CRNs. This way of description is followed in the next section.

To state our main results, 
we define the \textit{concentration integral} $\gev{\cdot}$ of a quantity $\bm{q}=(q_1,q_2,\dots,q_N)^\mathsf{T}$ as 
\begin{align}
    \gev{\bm{q}} := \sum_{i=1}^N q_i[\X_i]. 
\end{align}
A concentration integral coincides with an average if the weight is normalized. 
It might be possible to use not the concentration but the normalized concentration $[\X_i]/\sum_{i=1}^N[\X_i]$ or the mole fraction as the weight. 
However, usually both of them do not obey any tractable differential equations like the rate equation. 
Since we would like to consider time evolution, we use a concentration integral. 

We further define the chemical variance of a chemical potential $\gev{\Delta \bm{\mu}^2}$ as 
\begin{align}
    \gev{\Delta \bm{\mu}^2}:=
    \gev{(\bm{\mu}-\bm{\mu}^\eq)^2} = \sum_{i=1}^N (\mu_i-\mu_i^\eq)^2[\X_i], \label{cvcp}
\end{align}
where $(\bm{\mu}-\bm{\mu}^\eq)^2$ means the vector $\pqty{(\mu_i-\mu_i^\eq)^2}_{i=1,2,\dots,N}$. It is a variance-like quantity, but differs from a variance in two aspects. One is that the weight is not normalized, namely, it is defined by the concentration integral. The other is that what is subtracted from the chemical potential $\bm{\mu}$ is not an average but the equilibrium chemical potential $\bm{\mu}^\eq$.

One of our main results is the fact that the time derivative of the Gibbs free energy is bounded above by the product between the Fisher information and the chemical variance of a chemical potential. First, for closed CRNs, it is the inequality 
\begin{align}
    \abs{\frac{dG}{dt}}\leq \sqrt{\I(t)}\sqrt{\gev{\Delta \bm{\mu}^2}}, \label{mainineq}
\end{align}
where $\abs{\cdot}$ means the absolute value. 
This inequality gives an upper bound of the speed at which the Gibbs free energy decreases with the information geometric quantity, the Fisher information $\I(t)$. 
The inequality is rewritten equivalently as 
\begin{align}
    \frac{1}{\I(t)}\pqty{\frac{dG}{dt}}^2\leq \gev{\Delta \bm{\mu}^2}, \label{mainineq2}
\end{align}
which is similar to the Cram\'{e}r--Rao inequality Eq.~\eqref{crineq}. 
We define a function of a quantity $\bm{q}$, a reference value $\bar{\bm{q}}$, and the time $t$ as 
\begin{align}
    v_{\bm{q}}(t,\bar{\bm{q}}):=\sqrt{\I(t)}\sqrt{\gev{(\bm{q}-\bar{\bm{q}})^2}}, \label{speedfunction}
\end{align}
where $(\bm{q}-\bar{\bm{q}})^2=\pqty{(q_i-\bar{q}_i)^2}_{i=1,2,\dots,N}$, then Eq.~\eqref{mainineq} can be rewritten as 
\begin{align}
    \abs{\frac{dG}{dt}}\leq v_{\bm{\mu}}(t,\bm{\mu}^\eq). 
\end{align}

The proof of Eq.~\eqref{mainineq} is straightforward. From Eq.~\eqref{GibbsDiv}, we have 
\begin{align}
    \frac{dG}{dt}&=RT\sum_{i=1}^N\frac{d[\X_i]}{dt}\ln\frac{[\X_i]}{[\X_i]^\eq} \label{44}\\
    &=\sum_{i=1}^N\frac{d[\X_i]}{dt}(\mu_i-\mu_i^\eq) \label{45}, 
\end{align}
then, using the Cauchy--Schwarz inequality, the inequality is obtained as follows 
\begin{align}
    \abs{\frac{dG}{dt}}&=\abs{\sum_{i=1}^N\frac{1}{\sqrt{[\X_i]}}\frac{d[\X_i]}{dt}\sqrt{[\X_i]}(\mu_i-\mu_i^\eq)}\\
    &\leq \sqrt{\sum_{i=1}^N\frac{1}{[\X_i]}\pqty{\frac{d[\X_i]}{dt}}^2}
    \sqrt{\sum_{i=1}^N[\X_i](\mu_i-\mu_i^\eq)^2}. \label{pf}
\end{align}
Here we do not use any approximation or assumption of near-equilibrium, thus the speed limit holds far from equilibrium. 

The above discussion can be extended to subsystems. 
If we are concerned with some specific species $S=\{\X_i\}_{i\in A_S}\subset \{\X_1,\X_2,\dots,\X_N\}$, where $A_S$ is the index set of $S$, we can define the partial Gibbs free energy $G_S$ as 
\begin{align}
    G_S:=RT\sum_{i \in A_S}\pqty{[\X_i]\ln\frac{[\X_i]}{[\X_i]^\eq}-[\X_i]+[\X_i]^\eq}. 
\end{align}
The inequality Eq.~\eqref{mainineq} also holds for this partial Gibbs free energy 
\begin{align}
    \abs{\frac{dG_S}{dt}}\leq \sqrt{\I_S(t)}\sqrt{\gev{\Delta\bm{\mu}^2}_S}=:v_{\bm{\mu},S}(t,\bm{\mu}^\eq), \label{partialineq}
\end{align}
where 
\begin{align}
    \I_S(t)&:=\sum_{i \in A_S}\frac{1}{[\X_i]}\pqty{\frac{d[\X_i]}{dt}}^2\\
    \gev{\Delta\bm{\mu}^2}_S&:=\sum_{i \in A_S}(\mu_i-\mu_i^\eq)^2[\X_i]. 
\end{align}
The partial Gibbs free energy can show nontrivial behavior, e.g., oscillation, even in a closed CRN, while the Gibbs free energy of a total system decreases monotonically. Hence, this bound Eq.~\eqref{partialineq} also becomes a nontrivial one. 
The proof is almost the same as of Eq.~\eqref{mainineq}. 

Furthermore, the speed limit obtains in even open CRNs for the transformed Gibbs free energy $\G$. 
The time derivative of $\G$ is bounded above as the Gibbs free energy of closed systems was by the Fisher information and the deviation of chemical potential 
\begin{align}
    \abs{\frac{d\G}{dt}}\leq \sqrt{\I_{S_\X}(t)}\sqrt{\gev{(\bm{\mu}-\bm{\mu}^\ss)^2}_{S_\X}}, \label{speedlimitopen} 
\end{align}
where $S_\X=\{\X_1,\dots,\X_N\}$ and $(\bm{\mu}-\bm{\mu}^\ss)^2$ is the vector $\pqty{(\mu_i-\mu_i^\ss)^2}_{i=1,2,\dots,N+N'}$. Since the transformed Gibbs free energy is given by the $f$-divergence, one can prove the inequality Eq.~\eqref{speedlimitopen} easily. This inequality obtains in any open CRN which has a steady-state solution.

We comment an experimental importance of the speed limit. The inequality Eq.~\eqref{mainineq} can be written as follows 
\begin{align}
    \frac{1}{\sqrt{\gev{\Delta\bm{\mu}^2}}}\abs{\frac{dG}{dt}}\leq \frac{ds}{dt}. \label{frrlimit}
\end{align}
The denominator of the left-hand side is seen as the entire driving force of the CRN and the numerator the corresponding changing rate. 
Thus, the ratio itself can be interpreted as something like the transport coefficient in the linear response theory. 
The inequality Eq.~\eqref{frrlimit} shows that it is suppressed by the intrinsic speed on the stoichiometric compatibility class $ds/dt$. 
Hence if $ds/dt$ of a CRN is small, we can see that the CRN responds weakly to a change of chemical potentials. 
Since the intrinsic speed $ds/dt$ needs only the current concentration and the concentration's time derivative, it is experimentally obtained more easily than the transport coefficient like quantity that needs standard chemical potential or equilibrium concentration.
In addition, it is remarkable that the information-geometric speed $ds/dt$ can be defined and considered in CRNs with multistability and sustained oscillation, though these systems are out of the range of our study. 

\subsection{Generalized Cram\'{e}r--Rao inequality for chemical reaction networks}\label{result2}
\begin{figure}
    \begin{center}
        \includegraphics[width=\hsize]{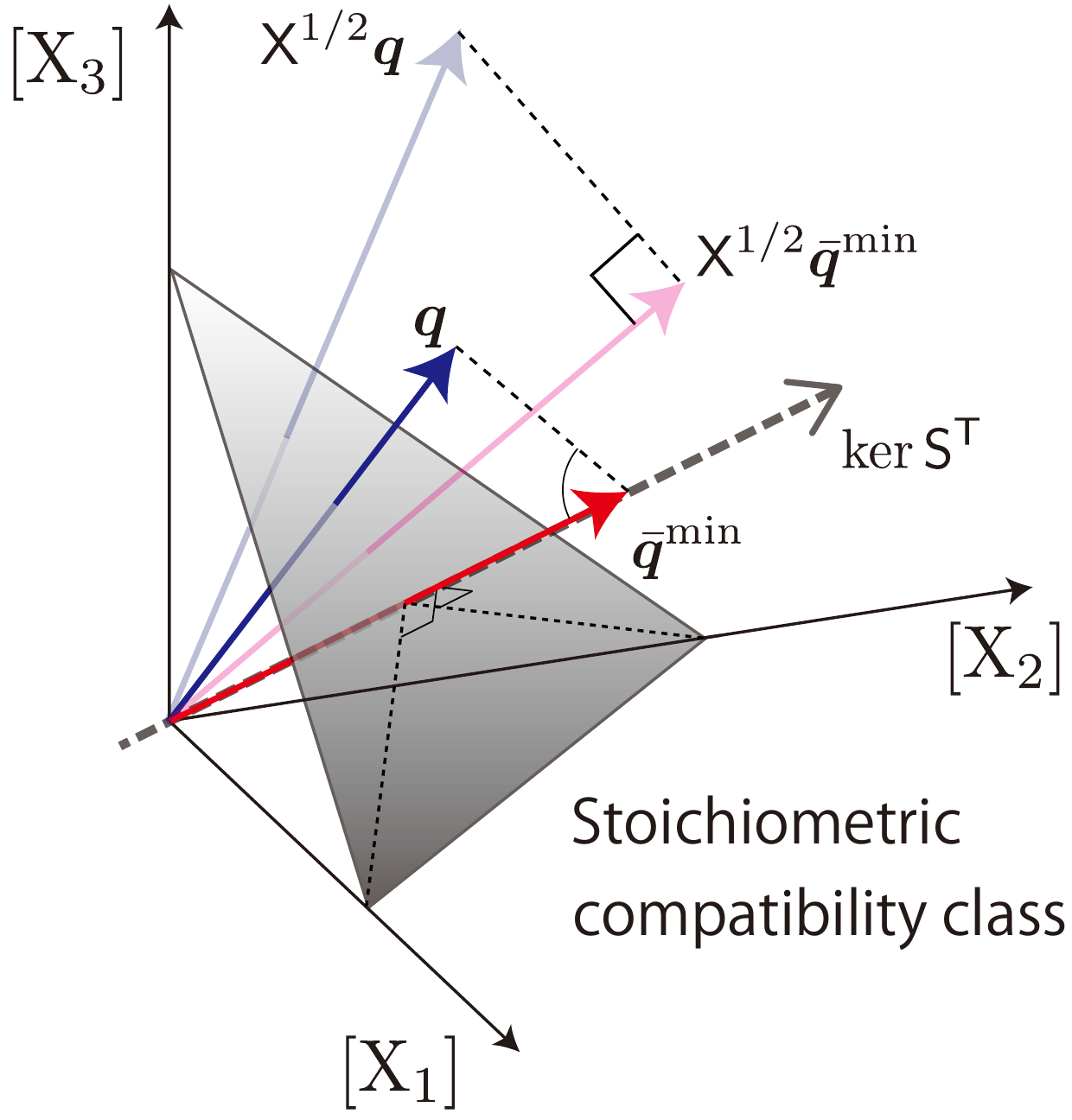}
        \caption{Geometric picture of $\bm{q}$ and other quantities appearing in Sec.\ref{result2} for $N=3$. 
        $\bar{\bm{q}}^\mathrm{\min}$, which makes the right hand sides of the inequalities Eq.~\eqref{gcr1} and Eq.~\eqref{gcr2} smallest, is obtained as a projection of $\bm{q}$ onto $\ker\st^\mathsf{T}$, which is perpendicular to the stoichiometric compatibility class. It is not the orthogonal projection in the Euclidean space. Instead, $\mathsf{X}^{1/2}\bm{q}$ is orthogonally projected to $\mathsf{X}^{1/2}\bar{\bm{q}}^\mathrm{min}$.}
        \label{okimochi}
    \end{center}
\end{figure}

We consider the result in the previous section further. 
If we use the original definition of the Gibbs free energy Eq.~\eqref{Gibbs1}, the time derivative of $G$ is given by 
\begin{align}
    \frac{dG}{dt}=\sum_{i=1}^N \frac{d[\X_i]}{dt}\mu_i, 
\end{align}
while in Eq.~\eqref{45} there is an additional term $-\sum_{i=1}^N(d[\X_i]/dt)\mu_i^\eq$. This is because $\bm{\mu}^\eq$ satisfies Eq.~\eqref{eqcond}, thus is orthogonal to $d[\bm{\X}]/dt=\st\bm{J}$. Hence, it is crucial for the speed limit Eq.~\eqref{mainineq} that $\bm{\mu}^\eq$ is a conservation law. This suggests that we can make use of conservation laws to evaluate the time derivative of a concentration integral. 

For example, letting $\bar{\bm{q}}$ be a conservation law, we can add $\bm{0}=-\bar{\bm{q}}\cdot(d[\bm{\X}]/dt)$ to the time derivative of $\gev{\bm{q}}$ 
\begin{align}
    \frac{d}{dt}\hspace{-2pt}\gev{\bm{q}}=\bm{q}\cdot \frac{d[\bm{\X}]}{dt}=(\bm{q}-\bar{\bm{q}})\cdot \frac{d[\bm{\X}]}{dt}. \label{gsl}
\end{align}
Then, we obtain the following inequality in the same way as the proof of Eq.~\eqref{mainineq} 
\begin{align}
    \abs{\frac{d}{dt}\hspace{-2pt}\gev{\bm{q}}}\leq \sqrt{\I(t)}\sqrt{\gev{(\bm{q}-\bar{\bm{q}})^2}}=v_{\bm{q}}(t,\bar{\bm{q}}),\label{gcr1}
\end{align}
or equivalently 
\begin{align}
    \frac{1}{\I(t)}\pqty{\frac{d}{dt}\hspace{-2pt}\gev{\bm{q}}}^2\leq \gev{(\bm{q}-\bar{\bm{q}})^2}.\label{gcr2}
\end{align}
We may call this inequality the generalized Cram\'{e}r--Rao inequality for CRNs. 
Eq.~\eqref{gcr1} reveals the fact that the Fisher information does not only acts as a speed limit on the Gibbs free energy, but also on general quantities. 

We can construct the $\bar{\bm{q}}\;(=:\bar{\bm{q}}^\mathrm{min})$ that minimizes $v_{\bm{q}}(t,\cdot)$. 
We introduce a diagonal matrix $\mathsf{X}$
\begin{align}
    \mathsf{X}_{ij}:=\delta_{ij}[\X_i],
\end{align}
then we can rewrite $\gev{(\bm{q}-\bar{\bm{q}})^2}$ as 
\begin{align}
    \sum_{i=1}^N(q_i-\bar{q}_i)^2[\X_i]&=\norm{\mathsf{X}^{1/2}(\bm{q}-\bar{\bm{q}})}^2\\
    &=\norm{\mathsf{X}^{1/2}\bm{q}-\mathsf{X}^{1/2}\bar{\bm{q}}}^2, \label{eq60}
\end{align}
where $\norm{\cdot}$ is the Euclidean norm. Since the linear space that $\mathsf{X}^{1/2}\bar{\bm{q}}$ belongs to is 
\begin{align}
    \mathsf{X}^{1/2}\ker\st^{\mathsf{T}}:=\{\mathsf{X}^{1/2}\bm{\ell}\mid\bm{\ell}\in\ker\st^{\mathsf{T}}\}, 
\end{align}
we see that Eq.~\eqref{eq60} is smallest when $\mathsf{X}^{1/2}\bar{\bm{q}}$ is the orthogonal projection of $\mathsf{X}^{1/2}\bm{q}$ onto $\mathsf{X}^{1/2}\ker\st^{\mathsf{T}}$ (see Fig.~\ref{okimochi}). 
Therefore, letting $\{\mathsf{X}^{1/2}\bm{\ell}^{\lambda}\}_{\lambda=1,2,\dots,\Lambda}$ be an orthonormal basis of $\mathsf{X}^{1/2}\ker\st^{\mathsf{T}}$, $\bar{\bm{q}}^\mathrm{min}$ is given by 
\begin{align}
    \bar{\bm{q}}^\mathrm{min}&=\mathsf{X}^{-1/2}\sum_{\lambda=1}^\Lambda (\mathsf{X}^{1/2}\bm{\ell}^{\lambda}\cdot\mathsf{X}^{1/2}\bm{q})\mathsf{X}^{1/2}\bm{\ell}^{\lambda}\\
    &=\sum_{\lambda=1}^\Lambda ((\bm{\ell}^{\lambda})^\mathsf{T}\mathsf{X}\bm{q})\bm{\ell}^{\lambda}. \label{62}
\end{align}

These results can be restricted to subsystems as Eq.~\eqref{partialineq}. For a subset of species $S$, we have 
\begin{align}
    \abs{\frac{d}{dt}\gev{\bm{q}}_S}&\leq v_{\bm{q},S}(t,\bar{\bm{q}})\\&:=\sqrt{\I_S(t)}\sqrt{\sum_{i \in A_S}(q_i-\bar{q}_i)^2[\X_i]}.
\end{align}
The right-hand side is minimized when $\bar{q}_i=\bar{q}^\mathrm{min}_i$ for all $i \in A_S$. 

Under near-equilibrium condition, it can be proved that $\bm{\mu}^\eq$ coincides with $\bar{\bm{\mu}}^\mathrm{min}$ within an error of 2nd order of the deviation $\Delta[\bm{\X}]:=[\bm{\X}]-[\bm{\X}]^\eq$. For any $\bm{\ell}\in\ker\st^\mathsf{T}$, 
\begin{align}
    &\pqty{\mathsf{X}^{1/2}(\bm{\mu}-\bm{\mu}^\eq)}\cdot \mathsf{X}^{1/2}\bm{\ell}\\
    &=RT\sum_{i=1}^N[\X_i]\ln\frac{[\X_i]}{[\X_i]^\eq}\ell_i\\
    &=RT\sum_{i=1}^N\ell_i([\X_i]-[\X_i]^\eq)+\mathcal{O}(\Delta[\bm{\X}]^2)\\
    &=RT\Delta\pqty{\bm{\ell}\cdot[\bm{\X}]}+\mathcal{O}(\Delta[\bm{\X}]^2), 
\end{align}
and $\bm{\ell}\cdot[\bm{\X}]$ does not change in time, therefore $\mathsf{X}^{1/2}(\bm{\mu}-\bm{\mu}^\eq)$ is approximately orthogonal to $\mathsf{X}^{1/2}\ker\st^{\mathsf{T}}$. This means that $\mathsf{X}^{1/2}\bm{\mu}$ is orthogonally projected to $\mathsf{X}^{1/2}\bm{\mu}^\eq\in \mathsf{X}^{1/2}\ker\st^{\mathsf{T}}$, therefore we see $\bm{\mu}^\eq\simeq\bar{\bm{\mu}}^\mathrm{min}$. Then the speed limit Eq.~\eqref{mainineq} is the tightest Cram\'{e}r--Rao bound under near-equilibrium conditions with an error of second order of the concentration deviation. 

Finally, let us extend the results to open CRNs. For a quantity $\bm{q}\in\mathbb{R}^{N+N'}$, a speed limit on the change of $\gev{\bm{q}}$ is given by
\begin{align}
    \frac{d}{dt}\hspace{-2pt}\gev{\bm{q}}
    =\sum_{i=1}^N q_i\sum_{\rho=1}^M\st_{i\rho}J_\rho 
\end{align}
because $d[\Y_j]/dt=0$. 
Thus it is bounded by quantities related to the internal species 
\begin{align}
    \abs{\frac{d}{dt}\hspace{-2pt}\gev{\bm{q}}}\leq 
    \sqrt{\I_{S_\X}(t)}\sqrt{\gev{(\bm{q}-\bar{\bm{q}})^2}_{S_\X}}, 
\end{align}
where $\bar{\bm{q}}$ is an element of $\ker(\st^\X)^\mathsf{T}\times\mathbb{R}^{N'}$. The generalized Cram\'{e}r--Rao bound for an open CRN is also obtained as 
\begin{align}
    \frac{1}{\I_{S_\X}(t)}\pqty{\frac{d}{dt}\hspace{-2pt}\gev{\bm{q}}}^2\leq \gev{(\bm{q}-\bar{\bm{q}})^2}_{S_\X}.
\end{align}

\subsection{Geometry of stoichiometric compatibility class}\label{sec_geoscc}
In addition to the above results, we consider an information-geometric aspect of stoichiometric compatibility classes. Both closed and open CRNs are considered in the same notation because the chemostatted species do not affect the geometrical structure. A concentration distribution is confined to a stoichiometric compatibility class. We can consider this confinement is a consequence of constraints $\bm{\ell}^\lambda\cdot[\bm{\X}]=L^\lambda(\mathrm{const.})$, besides the discussion in Sec.~\ref{SecClosed}. This is reminiscent of the situation of probability distributions where they are restricted by the normalization condition $\sum_{i}p_i=1$. In fact, this constraint is a key to deduce basic results in information geometry, such as \v{C}encov's theorem~\cite{amari2000methods,cencov1972statistical}. On the other hand, the more generic restrictions of chemical reactions need a nontrivial extension of information geometry. 

We indicate an information-geometric characterization of stoichiometric compatibility classes. Let $r_i:=\sqrt{[\X_i]}$ and $a_i^\lambda:=\sqrt{L^\lambda/|\ell_i^\lambda|}$. Note that $L^\lambda$ can be always non-negative by choosing $\bm{\ell}^\lambda$ properly. Then the metric with respect to $r_i$ is the Euclidean because 
\begin{align}
    ds^2=\sum_{i=1}^N\frac{1}{[\X_i]}d[\X_i]^2
    =\sum_{i=1}^N(2dr_i)^2. 
\end{align}
If $L^\lambda\neq0$, the constraint on concentration distributions becomes 
\begin{align}
    \sum_{i=1}^N \mathrm{sign}(\ell_i^\lambda)\pqty{\frac{r_i}{a_i^\lambda}}^2=1, 
\end{align}
which is an equation of a quadric surface, like an ellipsoid or a hyperboloid. 
If $L^\lambda=0$, the constraint represents a cone that contains the origin $[\bm{\X}]=\bm{0}$ as a specific case of quadric surfaces. 
This consideration shows that the stoichiometric compatibility class can be considered as the intersection between such quadric surfaces and the positive orthant. 

We also consider the geometry of reaction dynamics on a stoichiometric compatibility class. 
The length $\mathcal{L}$ of the path between the concentration at $t=0$, $[\bm{\X}]_{t=0}$, and at $t=\tau$, $[\bm{\X}]_{t=\tau}$ is defined as 
\begin{align}
    \mathcal{L}:=\int_0^\tau \frac{ds}{dt}dt
    =\int_0^\tau \sqrt{\sum_{i=1}^N\frac{1}{[\X_i]}\pqty{\frac{d[\X_i]}{dt}}^2}dt. 
\end{align}
From the speed limit on the Gibbs free energy Eq.~\eqref{mainineq}, the transport coefficient like quantity gives a lower bound to $\mathcal{L}$ 
\begin{align}
    \mathcal{L}\geq \int_0^\tau \frac{1}{\sqrt{\gev{\Delta\bm{\mu}^2}}}\abs{\frac{dG}{dt}} dt.
\end{align}
Moreover, applying the Cauchy--Schwarz inequality to $\mathcal{L}$, we obtain the inequality 
\begin{align}
    \mathcal{L}^2\leq 2\tau\mathcal{C}, \label{tradeoff0}
\end{align}
where $\mathcal{C}$ is called the thermodynamic cost and defined as
\begin{align}
    \mathcal{C}:=\frac{1}{2}\int_0^\tau dt\I(t). 
\end{align}
Here $1/2$ is a conventional coefficient like the $1/2$ in the action of a free particle in analytical mechanics.
As shown in Sec.~\ref{sec_va}, under near-equilibrium conditions, the Fisher information of a closed CRN is expressed as 
\begin{align}
    \I(t)=-\frac{1}{2R}\frac{d\sigma}{dt}
    +o(\Delta[\bm{\X}]^2), 
\end{align}
thus its integral is 
\begin{align}
    \int_0^\tau dt\I(t)= \frac{\sigma(t=0)-\sigma(t=\tau)}{2R}
    +o(\Delta[\bm{\X}]^2).
\end{align}
Since this quantity has the same dimension as reaction rates $J_\rho$ and $d[\X_i]/dt$, it may represent how fast the system relaxes in the time interval $\tau$ even far from equilibrium. Hence, we can interpret $\mathcal{C}$ as the mean relaxation rate. 
Then, we have more informative expression of the inequality Eq.~\eqref{tradeoff0} 
\begin{align}
    \tau\geq\frac{\mathcal{L}^2}{2\mathcal{C}}. \label{tradeoff1}
\end{align}
It shows a trade-off relation between time and the mean relaxation rate in CRNs. 
One might think that $\mathcal{L}$ and $\mathcal{C}$ have the same information because they differ just by the powers of integrands except for the coefficients. However, we cannot derive such a connection between $\mathcal{L}$ and the entropy production rate, so the two quantities represent different aspects of a system.

Apart from the dynamics, one can define the shortest distance between two distributions $[\bm{\X}],[\bm{\X'}]$ in the same stoichiometric compatibility class as
\begin{align}
    \mathcal{D}
    :=\underset{\bm{\gamma}}{\mathrm{inf}}
    \int_0^\tau \frac{ds}{dt}dt
    =\underset{\bm{\gamma}}{\mathrm{inf}}\int_0^\tau \sqrt{\sum_{i=1}^N\frac{1}{\gamma_i}\pqty{\frac{d\gamma_i}{dt}}^2}dt,
\end{align}
where the infimum is taken over the paths of concentration distribution that satisfy the initial and final conditions and are contained in the same stoichiometric compatibility class as $[\bm{\X}]$ and $[\bm{\X'}]$, namely  $\{\bm{\gamma}:[0,\tau]\to\mathcal{S}([\bm{\X}])\,|\,\bm{\gamma}(0)=[\bm{\X}],\bm{\gamma}(\tau)=[\bm{\X'}]\}$.
Since the shortest distance always serves as a lower bound of the length $\mathcal{L}$, another trade-off relation obtains 
\begin{align}
    \tau\geq\frac{\mathcal{D}^2}{2\mathcal{C}}.  \label{tradeoff2}
\end{align}
Although this inequality is a weaker bound than Eq.~\eqref{tradeoff1}, it gives a lower bound to the mean relaxation rate $\mathcal{C} \geq \mathcal{D}^2/2\tau$ which needs only information of the time interval and the initial and final distributions. 

\section{Examples}\label{sec_example}
Through three examples of CRN, we check our results, the speed limits, Eq.~\eqref{mainineq}, \eqref{partialineq}, the generalized Cram\'{e}r--Rao inequality, Eq.~\eqref{gcr1}, and the trade-off relations Eq.~\eqref{tradeoff1}, \eqref{tradeoff2}. 

\subsection{Speed limit in damped Brusselator}
\begin{figure}
    \begin{center}
        \includegraphics[width=\hsize]{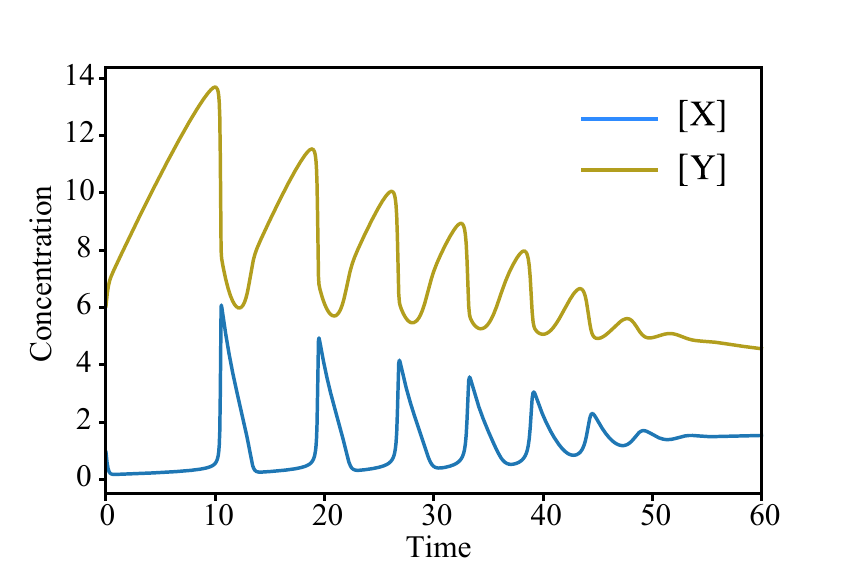}
        \caption{The time evolutions of $\X$ and $\Y$'s concentrations obtained by integrating Eq.~\eqref{brrateeq} with the parameters $k_1^+=1\times10^{-3},k_1^-=k_2^+=k_2^-=1,k_3^+=1\times10^{-2},k_3^-=1\times10^{-4}$, $[\X]_0=1,[\Y]_0=6,[\mathrm{A}]_0=[\mathrm{B}]_0=1\times10^{3}$. They oscillate, then relax to the equilibrium. }
        \label{br_concXY}
    \end{center}
\end{figure}

The first example is the Brusselator~\cite{prigogine1968symmetry,lefever1988brusselator}, which is a notable model of oscillating reactions such as the Belousov--Zhabotinsky reaction. We consider the following CRN 
\begin{equation}
    \begin{split}
    \mathrm{A}&\rightleftharpoons \mathrm{X}\\
    2\mathrm{X}+\mathrm{Y}&\rightleftharpoons 3\mathrm{X}\\
    \mathrm{X}+\mathrm{B}&\rightleftharpoons \mathrm{Y}+\mathrm{A}. 
    \end{split}
    \label{brreaction}
\end{equation}
Then the concentrations obey the rate equation below 
\begin{equation}
    \begin{split}
    \frac{d[\mathrm{X}]}{dt}&=J_1+J_2-J_3\\
    \frac{d[\mathrm{Y}]}{dt}&=-J_2+J_3\\
    \frac{d[\mathrm{A}]}{dt}&=-J_1+J_3\\
    \frac{d[\mathrm{B}]}{dt}&=-J_3,
    \end{split}\label{brrateeq}
\end{equation} 
where
\begin{align}
    J_1&=k_1^+[\mathrm{A}]-k_1^-[\mathrm{X}]\\
    J_2&=k_2^+[\mathrm{X}]^2[\mathrm{Y}]-k_2^-[\mathrm{X}]^3\\
    J_3&=k_3^+[\X][\mathrm{B}]-k_3^-[\Y][\mathrm{A}].
\end{align}
In a usual Brusselator model, the change of two species, $\X$ and $\Y$, are of interest, so the others are assumed to be constant because of the abundance. On the other hand, because we consider the system to be closed, we do not set $[\mathrm{A}]$ and $[\mathrm{B}]$ constant, but sufficiently large to observe the damped oscillation of $[\X]$ and $[\Y]$. 

\begin{figure}
    \begin{center}
        \includegraphics[width=\hsize]{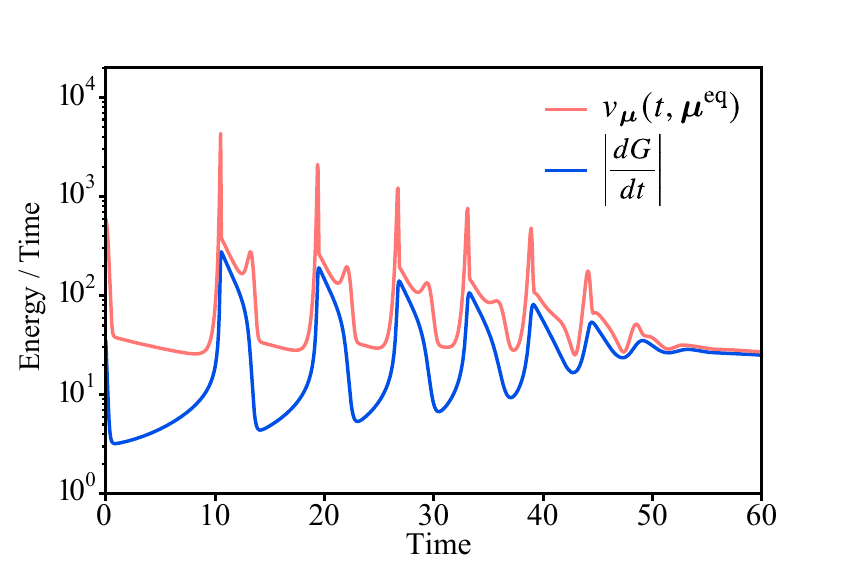}
        \caption{The speed limit Eq.~\eqref{mainineq} with respect to the reaction system~\eqref{brreaction}. The speed of the Gibbs free energy change cannot exceed $v_{\bm{\mu}}(t,\bm{\mu}^\eq)$. }
        \label{br_ineq}
    \end{center}
\end{figure}

\begin{figure}
    \begin{center}
        \includegraphics[width=\hsize]{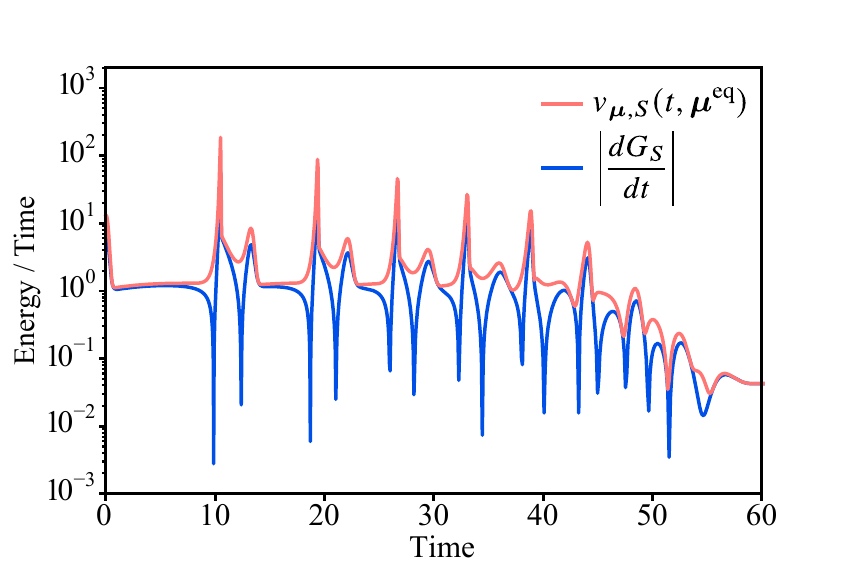}
        \caption{The speed limit Eq.~\eqref{mainineq} for a subset $S=\{\X,\Y\}$. The speed of the partial Gibbs free energy change also does not exceed the corresponding function $v_{\bm{\mu},S}(t,\bm{\mu}^\eq)$.}
        \label{br_ineq_partial}
    \end{center}
\end{figure}

While the Brusselator exhibits a damped oscillation of concentration as in Fig.~\ref{br_concXY}, 
$\abs{dG/dt}$ is suppressed by $v_{\bm{\mu}}(t,\bm{\mu}^\eq)$ as shown in Fig.~\ref{br_ineq}. The speed limit Eq.~\eqref{mainineq} is verified for the damped oscillatory CRN. Also the inequality for the partial Gibbs free energy Eq.~\eqref{partialineq} can be confirmed. Focusing on $S=\{\X,\Y\}$, we show its appearance in Fig.~\ref{br_ineq_partial}. 

There are situations where two curves are close, that is the inequality is tight. That occurs when the chemical potential changes exponentially. Because we used the Cauchy--Schwarz inequality to prove the speed limit in Eq.~\eqref{pf}, equality holds when there exists a constant $\alpha$ such that for all $i$, 
\begin{align}
    \quad\frac{1}{\sqrt{[\X_i]}}\frac{d[\X_i]}{dt}
    =\alpha\sqrt{[\X_i]}(\mu_i-\mu_i^\eq). 
\end{align}
It is equaivalent to 
\begin{align}
    \frac{d}{dt}(\mu_i-\mu_i^\eq)=\alpha'(\mu_i-\mu_i^\eq), 
\end{align}
where $\alpha'=RT\alpha$. 
Thus the equality holds when the deviation of the chemical potential $\mu_i-\mu_i^\eq$ is proportional to the exponential function $e^{\alpha' t}$. 
If the constant $\alpha'$ is negative, this condition means an exponential decay to the equilibrium. 
Under near-equilibrium conditions, the system is assumed to relax to the equilibrium exponentially. 
Hence, $\bm{\mu}^\eq$ should be the minimizer of the speed limit. 
We proved that $\bm{\mu}^\eq\simeq \bar{\bm{\mu}}^{\min}$ under near-equilibrium conditions in Sec.~\ref{result2}. 
Therefore, the equality condition is consistent with the previous discussion. 

\subsection{Generalized Cram\'{e}r--Rao inequality in a model where the total concentration does not conserve}
\begin{figure}
    \centering
    \includegraphics[width=\hsize]{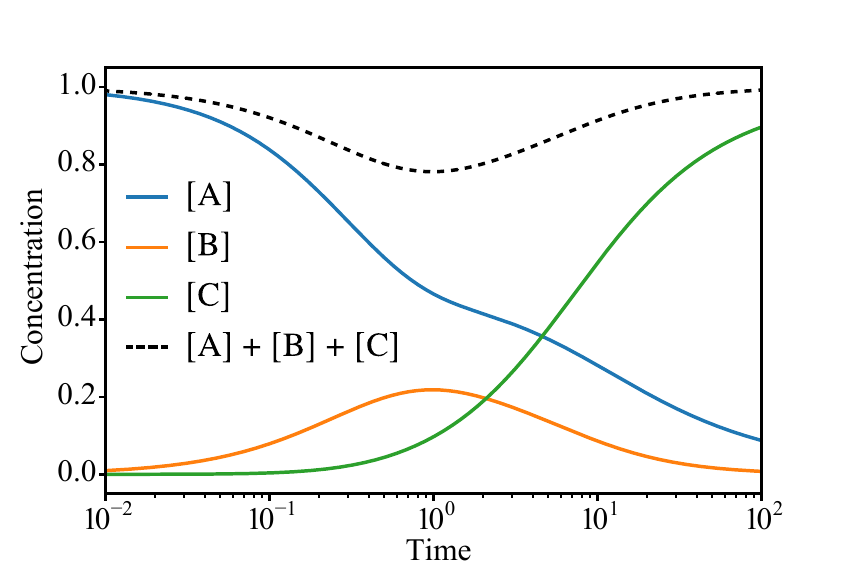}
    \caption{The time evolution of the concentrations in the CRN Eq.~\eqref{model2} calculated with the parameters $k_1^+=k_1^-=k_2^+=1,k_2^-=1\times10^{-5}$, $[\mathrm{A}]_0=1,[\mathrm{B}]_0=[\mathrm{C}]_0=1\times10^{-5}$. The total concentration $[\mathrm{A}]+[\mathrm{B}]+[\mathrm{C}]$ (dashed line) is not constant. }
    \label{model2_conc}
\end{figure}

We confirm the validity of the generalized Cram\'{e}r--Rao inequality Eq.~\eqref{gcr1} furthermore by observing the following CRN. 
\begin{equation}
    \begin{split}
        2\mathrm{A}&\rightleftharpoons\mathrm{B}\\
        \mathrm{A}+\mathrm{B}&\rightleftharpoons \mathrm{B}+\mathrm{C}
    \end{split}
    \label{model2}
\end{equation}
This CRN is simple but sufficient to break the conservation of total concentration as shown in Fig.~\ref{model2_conc}. Note that the preceding Brusselator model looks complicated but preserves the total concentration, $[\X]+[\Y]+[\mathrm{A}]+[\mathrm{B}]$, thus the concentration can be normalized by dividing this constant.

The stoichiometric matrix is 
\begin{align}
    \st=\pmqty{-2&-1 \\ 1&0 \\ 0&1}, 
\end{align}
so that the conservation law is only 
\begin{align}
    \bm{\ell}=\pmqty{1 \\ 2 \\ 1}
\end{align}
up to a scale factor. Therefore $[\mathrm{A}]+2[\mathrm{B}]+[\mathrm{C}]=:L$ becomes a constant instead of the total concentration. 

\begin{figure}
    \centering
    \includegraphics[width=\hsize]{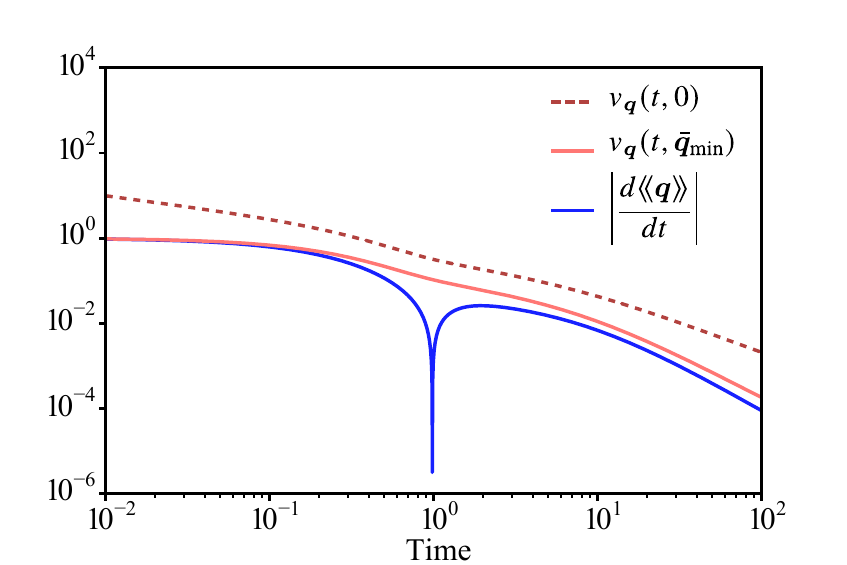}
    \caption{Bounds on the changing rate of the total concentration $\gev{\bm{q}}$. The bound given by $\bar{\bm{q}}^\mathrm{min}$ is much tighter than that given by $\bar{\bm{q}}=\bm{0}$. }
    \label{gcr_fig}
\end{figure}

If we set $\bm{q}=(1,1,1)^\mathsf{T}$, $\gev{\bm{q}}$ is the total concentration 
\begin{align}
    \gev{\bm{q}}=[\mathrm{A}]+[\mathrm{B}]+[\mathrm{C}]. 
\end{align}
The projection $\bar{\bm{q}}^\mathrm{min}$ is given by 
\begin{align}
    \bar{\bm{q}}^\mathrm{min}=\frac{\bm{\ell}^\mathsf{T}\mathsf{X}\bm{q}}{\bm{\ell}^\mathsf{T}\mathsf{X}\bm{\ell}}\bm{\ell}
    =\frac{L}{[\mathrm{A}]+4[\mathrm{B}]+[\mathrm{C}]}\bm{\ell},  
\end{align}
where a denominator $\bm{\ell}^\mathsf{T}\mathsf{X}\bm{\ell}$ appears unlike Eq.~\eqref{62} because $\mathsf{X}^{1/2}\bm{\ell}$ is not a unit vector here. 
Then the change of the total concentration is bounded as in Fig.~\ref{gcr_fig}. To compare with the tightest bound $v_{\bm{q}}(t,\bar{\bm{q}}^\mathrm{min})$, the bound given by the trivial conservation law $\bm{\bar{q}}=\bm{0}$, $v_{\bm{q}}(t,0)$, is presented together. 

\subsection{Trade-off relations on stoichiometric compatibility class}
\begin{figure}
    \begin{center}
        \includegraphics[width=\hsize]{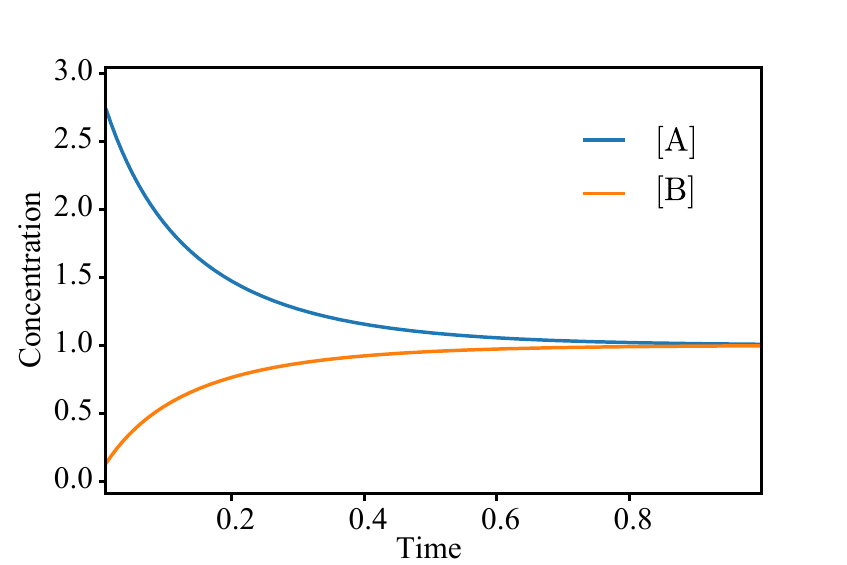}
        \caption{The time evolution of the concentrations in the association reaction. The parameters are set to $k^+=k^-=1$, $[\mathrm{A}]_0=2.9$ and $[\mathrm{B}]_0=0.05$. }
        \label{conc_ex3}
    \end{center}
\end{figure}
\begin{figure}
    \begin{center}
        \includegraphics[width=\hsize]{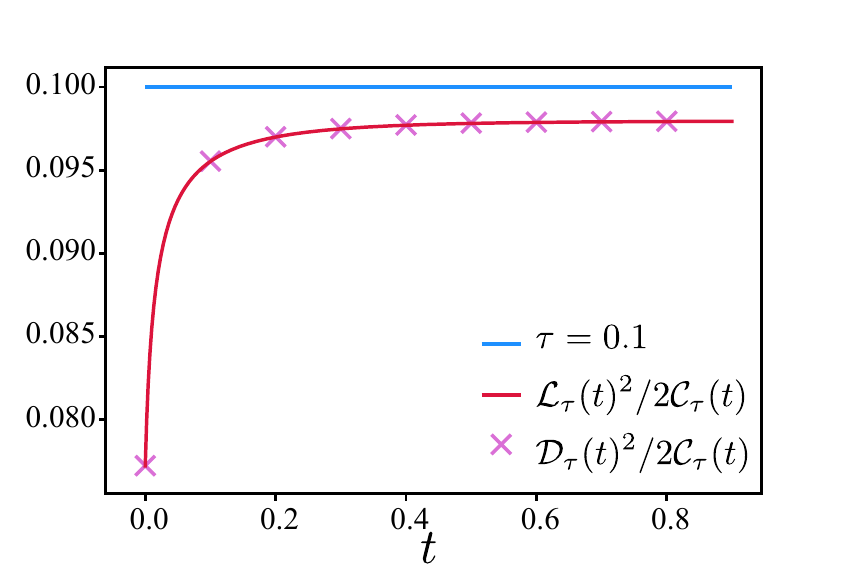}
        \caption{The trade-off relations Eq.~\eqref{tradeoff1} and Eq.~\eqref{tradeoff2} are verified for the CRN Eq.~\eqref{crnex3}. For the fixed $\tau=0.1$, $\mathcal{L}_\tau^2/2\mathcal{C}_\tau$ and $\mathcal{D}_\tau^2/2\mathcal{C}_\tau$ give a lower bound. Because the concentration change is monotonic, $\mathcal{L}_\tau$ and $\mathcal{D}_\tau$ give the same bound. 
        }
        \label{fig_tradeoff}
    \end{center}
\end{figure}

We illustrate our discussion in Sec.~\ref{sec_geoscc} by considering the following association reaction
\begin{align}
    2\mathrm{A}\rightleftharpoons\mathrm{B}.
    \label{crnex3}
\end{align}
For this CRN, shown in Fig.~\ref{conc_ex3}, we can obtain the distance $\mathcal{D}$ analytically. 
We denote the conserved quantity $[\mathrm{A}]+2[\mathrm{B}]$ as $L$. For new variables 
$r_1=2\sqrt{[\mathrm{A}]}$, $r_2=2\sqrt{[\mathrm{B}]}$, we can write the metric and the constraint as 
\begin{align}
    &ds^2=(dr_1)^2+(dr_2)^2\\
    &\qty(\frac{r_1}{2\sqrt{L}})^2+\qty(\frac{r_2}{\sqrt{2L}})^2=1.
\end{align}
Because this constraint represents an elliptic, the coordinate can be parametrized by a parameter $\theta$ as $(r_1,r_2)_{t=0}=(2\sqrt{L}\cos\theta_1,\sqrt{2L}\sin\theta_1)$, $(r_1,r_2)_{t=\tau}=(2\sqrt{L}\cos\theta_2,\sqrt{2L}\sin\theta_2)$. Then the distance is obtained as 
\begin{align}
    \mathcal{D}&=\abs{\int_{\theta_1}^{\theta_2}d\theta\sqrt{\pqty{\frac{dr_1}{d\theta}}^2+\pqty{\frac{dr_2}{d\theta}}^2}}\\
    &=2\sqrt{L}\abs{\int_{\theta_1}^{\theta_2}d\theta
    \sqrt{1-\frac{1}{2}\cos^2\theta}}\\
    &=2\sqrt{L}\abs{E\qty(\theta_2;\frac{1}{\sqrt{2}})-E\qty(\theta_1;\frac{1}{\sqrt{2}})},
\end{align}
where $E(x;k):=\int_0^x d\theta \sqrt{1-k^2\cos^2\theta}$ is the incomplete elliptic integral of the second kind. 

We introduce the following notation to check the trade-off relation for time $t$ and time interval $\tau$
\begin{align}
    \mathcal{L}_\tau(t)&:=\int_t^{t+\tau}dt'
    \frac{ds}{dt'}\\
    \mathcal{C}_\tau(t)&:=\frac{1}{2}\int_t^{t+\tau}dt'\I(t')\\
    \mathcal{D}_\tau(t)&:=\underset{\bm{\gamma}}{\mathrm{inf}}\int_t^{t+\tau} dt'\frac{ds}{dt'},
\end{align}
where the infimum is taken over $\bm{\gamma}$'s that satisfy $\bm{\gamma}(t)=[\bm{\X}]_t$ and $\bm{\gamma}(t+\tau)=[\bm{\X}]_{t+\tau}$ and are contained the same stoichiometric compatibility class as $[\bm{\X}]_t$. 
For the fixed time interval $\tau=0.1$, the trade-off relations are shown in Fig.~\ref{fig_tradeoff}.
Since the relaxation is monotonic, the length
$\mathcal{L}$ and the distance $\mathcal{D}$ should coincide with each other. In Fig.~\ref{fig_tradeoff}, $\mathcal{L}_\tau(t)^2/2\mathcal{C}_\tau(t)$ and $\mathcal{D}_\tau(t)^2/2\mathcal{C}_\tau(t)$ are actually the same. 

\section{Conclusion}\label{sec_conclusion}
We have studied thermodynamics of chemical reaction networks in terms of information geometry. 
We have revealed that geometrical structure and the Fisher information can be obtained in CRNs by using information geometry. 
Then we have derived speed limits in CRNs, e.g., Eq.~\eqref{mainineq}, \eqref{speedlimitopen}, and \eqref{gcr1}. Our results are not restricted to near-equilibrium conditions but hold even if the CRN is open or far from equilibrium. This broad range of application shows the universality of the speed limit. 
It has been shown that the speed limit can be interpreted as a generalization of the Cram\'{e}r--Rao inequality Eq.~\eqref{gcr2} outside probability spaces. These results are mainly based on the form of the Gibbs free energy that includes the $f$-divergence $D([\bm{\X}]\|[\bm{\X}]^\eq)$ and the conservation quantities dwelling in a CRN. 
We have further obtained a trade-off relation Eq.~\eqref{tradeoff1} between time and speed in CRNs examining the geometry of stoichiometric compatibility classes.

Our study provides a new perspective on chemical thermodynamics in terms of information geometry. 
It offers a framework to analyze the thermodynamic profile of biological systems. The use of the information-geometric measures such as the intrinsic speed $ds/dt$ or the length of reactions $\mathcal{L}$ would bring a new perspective to informatic aspects of biology. Our results can be used if one can obtain the concentration distribution, so the range of application would be wide.

There is a more theoretical question. Though the information geometry of chemical thermodynamics is brought by the $f$-divergence, one can ask whether there are more fundamental reasons why information geometry is applicable to CRNs. CRNs are just a kind of dynamical systems and have nothing to do with probability in its formulation. The fact that however they can be studied by using information geometry, which is usually useful in probability theory, would be a clue to investigate the link between various dynamical systems in nature and information. 

\begin{acknowledgements}
We thank Keita Ashida, Kiyoshi Kanazawa, Andreas Dechant, Takahiro Sagawa and Shin-ichi Sasa for fruitful discussions. 
S. I. is supported by JSPS KAKENHI Grant No. 19H05796 and JST Presto Grant No. JPMJPR18M2. 
\end{acknowledgements}

\appendix

\end{document}